\documentclass[11pt]{article}
\usepackage[margin=1in]{geometry}

\usepackage{amsmath,amssymb}
\usepackage{tikz-cd}
\usepackage{multicol}
\usepackage[left]{lineno}
\usepackage[hidelinks]{hyperref}
\usepackage{xurl}
\usepackage{graphicx}
\usepackage[font=small]{caption}
\usepackage{booktabs}
\usepackage{enumitem}
\DeclareCaptionFormat{linenumbered}{\internallinenumbers#1#2#3}
\title{\textbf{Skewed weak and Pareto-tailed strong interactions accompany community complexity}}
\author{%
\begin{tabular}{c}
Takuya Hojo$^{1\dagger}$\hspace{1.2em}Taikou Arakaki$^{2,3}$\hspace{1.2em}Koichi Fujimoto$^{3\dagger}$\\[0.4em]
\parbox{0.95\textwidth}{\centering
{\small $^{1}$Department of Biological Sciences, Graduate School of Science, The University of Tokyo, Faculty of Science Bld. 2, Hongo Campus, 7-3-1 Hongo, Bunkyo-ku, Tokyo 113-0033, Japan}\\
{\small $^{2}$Department of Biological Sciences, Graduate School of Science, Osaka University, 1-1 Machikaneyama-cho, Toyonaka, Osaka 560-0043, Japan}\\
{\small $^{3}$Program of Mathematical and Life Sciences, Graduate School of Integrated Sciences for Life, Hiroshima University, 1-3-1 Kagamiyama, Higashi-Hiroshima, Hiroshima 739-8526, Japan}\\[0.35em]
{\small $^{\dagger}$Corresponding authors: \href{mailto:hojo@ubi.s.u-tokyo.ac.jp}{hojo@ubi.s.u-tokyo.ac.jp}; \href{mailto:kfjmt@hiroshima-u.ac.jp}{kfjmt@hiroshima-u.ac.jp}}
}
\end{tabular}
}
\date{May 17, 2026}

\newcommand{\supportinggraphic}[2][]{%
\IfFileExists{#2}{%
\includegraphics[#1]{#2}%
}{%
\fbox{\parbox[c][0.35\textheight][c]{0.9\textwidth}{\centering Missing figure file: \texttt{\detokenize{#2}}}}%
}%
}

\begin{document}
\maketitle


\section{Abstract}

Ecological communities are often characterized by many weak and few strong interspecific interactions, yet their quantitative structure, generative basis, and links to community-level properties remain poorly understood. Using two empirical datasets of plant--animal networks, we show that both trophic and mutualistic interaction strengths distribute skewed weak and Pareto-strong tails (SWAPS), as quantified by positive skewness and extreme value theory, respectively. We further find that interaction strengths are taxon-specific and largely constrained within taxa. In community assembly simulations based on a generalized Lotka--Volterra model, this taxonomic conservatism, together with multiple interaction types beyond trophic and mutualistic ones, is required for the emergence of SWAPS distribution. Notably, SWAPS distribution emerges not only at the species level but also across lineages, and its emergence accompanies increases in community complexity. Together, these results identify SWAPS distribution as a previously unrecognized interaction signature of ecological communities and provide a new perspective on the organization of community-level properties.

\section{Introduction}

Ecological communities are complex systems composed of taxonomically distant and close species that have diversified through evolution [1, 2, 3, 4, 5]. These species engage in multiple types of direct interspecific interactions, including predator-prey, mutualism, and parasitism [6] which constitute the interaction network within the community [2, 3, 4, 5, 7]. Elucidating the network structure , i.e., topology and interaction strength, could help us understand how species diversify and how community complexity increases [8, 9, 10].

While topological structure of interaction networks, i.e., degree distribution and trophic layering, has been extensively studied [3, 11, 12], it remains little investigated on the structure of interaction strength (IS). IS represents the ecological impact of one species on another, and is considered a key feature of the ecological network structures [10, 13, 14]. The distribution of the ISs within a community reflects its overall structure and function. In communities consisting solely of trophic interactions (i.e., food webs), IS distributions are typically skewed towards weak interactions [15, 16, 17], hereafter referred to as skewed weak interactions (SW). Moreover, some studies have pointed out that these distributions also encompass a significant proportion of strong interactions [18, 19], suggesting that ISs are heavy-tailed as well as SW. SW has been suggested to improve the community stability [20, 21, 22, 23, 24]. Regarding interaction types other than predator-prey, a few studies have shown that mutualistic interactions also follow SW patterns [25, 26]. Additionally, the IS distributions, inferred from population time series data of multiple fish species within communities using empirical dynamic modeling and related nonlinear time-series methods, were also broadly consistent with SW, although heavy tails were not explicitly examined in those studies [27, 28]. Thus, it remains unclear whether mutualistic interactions exhibit heavy tails and whether such multiple interaction types coexisting within a single community follow SW and heavy tail. A more systematic and quantitative assessment of IS could reveal a shared principle across multiple types of interspecific interaction.

Given the conserved tendency that closely related species interact with the same partner taxa in mutualistic and antagonistic networks [5, 29, 30], an unresolved question is whether their interaction strengths are also conserved [31, 32]. For instance, are interactions between one pair of taxonomic groups (e.g., taxa A and C in Fig. 1A upper) consistently strong, while those between other taxonomic pairs (taxa A and D; B and C; B and D) are consistently weak? However, this taxonomic conservatism and specificity of IS remains little investigated quantitatively. If ISs are conserved within taxonomic combinations, the weak and strong nature of the IS distribution may persist even after averaging ISs at higher taxonomic levels. Therefore, examining how strong and weak interactions are distributed within and across taxonomic groups would provide clues to the emergence process of SW and heavy tail distribution.

While it is extremely difficult to empirically study how SW and heavy tail distribution emerges, mathematical modeling offers a powerful alternative. The Generalized Lotka–Volterra (GLV) model, for example, has been widely used to simulate complex interaction networks [24, 33, 34]. Since interspecies interaction strength in GLV model represents the effect of a species on the per-capita growth of other species, it has been considered that this interaction strength (coefficient) reflects the relative magnitude of empirically measured IS [13, 35, 36], while the IS of GLV model differs from the measured IS in units, estimation methods, and biological interpretation [37].  Community assembly in such model has contributed to our understanding of how species immigration, environmental filtering, and evolutionary history shape community structure, by incorporating processes such as immigration [38, 39, 40] and speciation—also referred to as phylogenetic evolution [41, 42, 43, 44, 45]. Notably, several studies employing such models have demonstrated the emergence of multispecies coexistence through community assembly involving immigration and speciation [46, 47]. Importantly, these theoretical studies, including May’s theory known for the representative study about community complexity, have assumed that interspecific interactions (or the nonzero elements of the Jacobian matrix) follow the Gaussian distribution, which lacks the observed SW and heavy tail distribution [20, 48]. Therefore, exploring the generative principle of SW and heavy tail distribution through community assembly simulations would offer new perspectives to increasing community diversity and complexity.

The primary aims of this study are to examine whether predator-prey and mutualistic interactions follow SW and heavy tail distribution in the multispecies coexisting community, and to reveal the generative principle of SW and heavy tail distribution and the correlation of SW and heavy tail distribution emergence with community complexity. To this end, we quantitatively analyze the open data [49, 50] and show heavy tail as well as SW with taxonomic specificity and conservatism. The taxonomic conservatism and specificity further derived SW and heavy tail distribution averaged at each taxonomic group. Our assembly simulations of the GLV model revealed that both the taxonomic conservatism on IS and multiple interaction types necessitate the emergence of SW and heavy tail distribution. Moreover, SW and heavy tail distribution emergence positively correlates with the increase of community complexity.

\section{Materials and Methods}

\subsection{Community data}

We analyzed publicly available datasets of interspecies interactions comprising many species and multiple types of interaction, including mutualistic and mixed trophic--mutualistic communities, to examine IS distributions [49, 50]. Within the agroecosystem, we examined IS distributions for mutualistic and trophic interactions, for which sufficient data were available among the three recorded types of interspecies interaction: mutualistic, trophic, and parasitic.

\subsection{Investigation of interspecies interactions in relation to taxa}

To examine whether IS is specific to orders or families, we gathered each species' taxonomic information in the data from open-source databases such as the Global Biodiversity Information Facility (GBIF). In this analysis, we excluded network nodes that were not resolved to the species level, including those identified only at higher taxonomic levels such as genus, family, or order. Taxonomic conservatism was evaluated for clade pairs with more than four interactions using the coefficient of variation (CV) of ISs from clade $X$ to clade $Y$ (Fig.~2B), defined as
\begin{linenomath*}
\[
\mathrm{CV}_{Y X} = \frac{\sigma_{Y X}}{\mu_{Y X}},
\]
\end{linenomath*}
where $\mu_{Y X}$ and $\sigma_{Y X}$ indicate the mean and standard deviation of IS from clade $X$ to clade $Y$, respectively. As a null model for the clade pairs, we generated a shuffled network that preserved network topology and the IS distribution while randomly swapping IS across links over the entire network.

\subsection{Generalized Lotka--Volterra model}

We simulated the temporal change in biomass or abundance $x_i$ of multiple species ($i = 1, \ldots, M$) using a GLV-type linear interaction model [41, 51, 52]:
\begin{linenomath*}
\begin{equation}
\frac{d x_i}{d t} = f_i(x) = \left(r_i + \sum_{j=1}^{M} \alpha_{i j} x_j\right)x_i,
\tag{Eq. 1}
\end{equation}
\end{linenomath*}
where $r_i$ and $\alpha_{i j}$ denote the intrinsic reproductive rate and the $(i,j)$ entry of the interaction matrix $\alpha$, representing the effect of species $j$ on the per-capita growth of species $i$, respectively. The summation term aggregates influences from all interacting species in the community, allowing the net growth of each species to be determined by its intrinsic growth and the combined effects of interspecific interactions.

For each off-diagonal pair $(i,j)$, we classified interaction type based on the signs of $\alpha_{i j}$ and $\alpha_{j i}$: commensalism ($+/0$) or amensalism ($-/0$) when one entry was zero, interference-type competition ($-/-$), mutualism ($+/+$), or predator--prey/parasitism ($-/+$). In the interaction of prey $i$ and predator $j$, $\alpha_{i j} < 0$ and $\alpha_{j i} > 0$, we imposed a directional constraint $\alpha_{j i} = \mathrm{eff}\, |\alpha_{i j}|$, where $\mathrm{eff}$ is the energy absorption efficiency and was fixed to 0.7 in the present simulations. Interaction strength was defined as the average of the absolute values of nonzero entries, i.e., $(|\alpha_{i j}| + |\alpha_{j i}|)/2$ for competition, mutualism, and predator--prey/parasitism, and $|\alpha_{i j}|$ for commensalism and amensalism.

\subsection{Modeling community assembly}

We augmented the GLV-based community model to include two demographic processes for each generation of community assembly: (i) extinction and (ii) immigration of phylogenetically close or distant relatives of resident species.

\paragraph{Extinction.}
After numerically integrating the GLV dynamics up to time $T_{\mathrm{end}}$, we computed the mean biomass of each species over the final time window of length 2,500 to reduce the influence of temporal biomass fluctuations. A species was regarded as extinct when its mean biomass over this final time window was smaller than a threshold $\varepsilon = 1.0 \times 10^{-8}$. Once identified as extinct, the species was removed from the community together with the parameters $r_i$ and $\alpha_{i j}$ involving that species.

\paragraph{Immigration of phylogenetically close species.}
We introduced a new species $\rho'$ phylogenetically close to one resident species $\rho \in \{1,2,\ldots,M\}$, which was randomly selected from the community. We defined the intrinsic reproductive rate $r_{\rho'}$ of the new species as
\begin{linenomath*}
\[
r_{\rho'} = r_{\rho} + \eta_{\rho},
\]
\end{linenomath*}
where $\eta_{\rho}$ follows a Gaussian distribution with mean 0 and standard deviation 0.1, i.e., $\mathcal{N}(0,0.1)$. For each linked pair, outgoing and incoming interactions were mutated independently with probability 15\% as $\alpha_{\rho' j} = \alpha_{\rho j} + \eta_{\rho j}$ and $\alpha_{i \rho'} = \alpha_{i \rho} + \eta_{i \rho'}$, respectively, subject to sign preservation and the constraint $\alpha_{j \rho'} = \mathrm{eff}\, |\alpha_{\rho' j}|$ for trophic interactions, where $\eta_{\rho j}$ and $\eta_{i \rho'}$ also follow $\mathcal{N}(0,0.1)$. When an interaction was newly introduced, magnitudes were initialized from a uniform distribution on $[0,1)$ with prescribed signs. Additionally, we assumed that addition or deletion of an interaction occurred with probability 5\%, where added interactions were initialized in the same manner. We did not introduce interference competition between closely related species $\rho$ and $\rho'$, following Cahill et al.~(2008). The initial population of newly added species $\rho'$ was set at 10\% of that of species $\rho$ after extinction was completed. We defined a lineage as a group comprising species diverging from the same ancestral species, similarly to the taxonomic groups used in the data analysis.

\paragraph{Immigration of phylogenetically distant species.}
We introduced a new species $\rho'$, independent of resident species, into the community. The presence or absence of species interactions between species $\rho'$ and $j$ ($j = 1,2,\ldots,M$) was determined with probability $P_j$, defined as the number of interspecies interactions possessed by species $j$ divided by the total number of interspecies interactions in the entire community. This reflects the preferential attachment mechanism necessary for forming the characteristic network structure, i.e., scale-free networks, observed in communities [54]. The interspecific interactions $\alpha_{\rho' j}$ and $\alpha_{j \rho'}$ and the intrinsic growth rate $r_{\rho'}$ were determined in the same manner as the initial setting in the assembly simulation described below. The initial population size of newly added species $\rho'$ was set to 0.1.

\subsection{Numerical simulation}

The community assembly was initialized with $M = 20$ species. The connectance $C$ of the interaction matrix $\alpha$, defined later as the fraction of interacting species pairs among all possible ones, was initially set to 0.1. The diagonal elements $\alpha_{ii}$ were initially set to $-1$ for all species, representing intraspecific density dependence [43, 44, 55, 56], and were specified independently of connectance $C$. The nonzero interaction coefficients $\alpha_{ij}$ ($i \neq j$) and intrinsic growth rates $r_i$ were randomly drawn from a uniform distribution on $[-1,1]$. The initial population of each species was set to $x_i = 0.1$.

Under these settings, we numerically integrated the GLV dynamics up to $T_{\mathrm{end}} = 1.0 \times 10^{4}$. Because community assembly often generates dynamics with strongly separated timescales, including fast transients and slow relaxations, explicit solvers could become numerically unstable or require prohibitively small time steps. We therefore used \texttt{RadauIIA5()}, an implicit fifth-order Runge--Kutta solver of the Radau IIA family implemented in DifferentialEquations.jl (Julia), which is well suited for such multiscale dynamics.

Community assembly was performed for 1000 generations for each assembly condition: (I) immigration of phylogenetically distant species only (Section 2.4, hereafter the condition without taxonomic conservatism), and (II) immigration of taxonomically close species (Section 2.4), except that one in every five immigration events involved a distantly related species (hereafter the condition with taxonomic conservatism). In each condition, we calculated community assemblies with fifty independent initial conditions. When simulating communities split off by network division, the community with the greatest species richness was investigated. If community assembly simulations did not successfully integrate over the predefined time window, as indicated by an unsuccessful solver return code and/or pathological time stepping, the proposed species immigration was rejected and the immigration process was retried. We repeated this up to 300 times per generation until the simulation was successfully integrated; otherwise, immigration was terminated and the simulation proceeded to the next generation.

\subsection{Statistical analysis}

\paragraph{Skewness test.}
We evaluated the IS distribution by the first to third moments:
\begin{linenomath*}
\[
\mu = \frac{1}{n}\sum_{i=1}^{n} y_i,
\]
\end{linenomath*}
\begin{linenomath*}
\[
\sigma^2 = \frac{1}{n}\sum_{i=1}^{n} y_i^2 - \mu^2,
\]
\end{linenomath*}
\begin{linenomath*}
\[
\beta = \frac{n}{(n-1)(n-2)}\sum_{i=1}^{n}\left(\frac{y_i-\mu}{\sigma}\right)^3,
\]
\end{linenomath*}
where $n$ is the number of interspecies interactions. The third moment was normalized by the second moment. We used the skewness test statistic following Jarque and Bera (1987):
\begin{linenomath*}
\[
Z_{\mathrm{skew}} = \frac{\beta}{\mathrm{SE}_{\mathrm{skew}}}, \qquad \mathrm{SE}_{\mathrm{skew}} = \sqrt{\frac{6}{n}}.
\]
\end{linenomath*}
Under the null expectation of no skewness, we evaluated $Z_{\mathrm{skew}}$ using a right-tailed one-sided test against the Gaussian distribution, because our hypothesis specifically concerned positive skewness, corresponding to a distribution skewed toward weak interactions with a heavy right tail. Statistical significance was defined as $Z_{\mathrm{skew}} > 1.645$, equivalent to $p < 0.05$ for a one-sided Gaussian test.

\paragraph{Quantification of tail weight based on the peak-over-threshold theorem.}
To quantify tail heaviness, we used a peaks-over-threshold approach grounded in extreme value theory and fitted a generalized Pareto distribution to threshold exceedances [58, 59, 60]. In the generalized Pareto family, the shape parameter, or Pareto exponent, $\xi$ governs tail decay, and the case $\xi = 0$ corresponds to the exponential distribution [61]; thus, the exponential model serves as a natural reference for an exponential-type tail. We compared the exponential model ($\xi = 0$) with the generalized Pareto model using log-likelihoods, and we assessed uncertainty in $\xi$ using a bootstrap procedure to enable confidence-interval-based classification. Thresholds were selected reproducibly under constraints on exceedance count and exceedance fraction, allowing consistent evaluation across distributions with different sample sizes (see Supporting Information Method S1 for details).

\paragraph{Mann--Whitney U test.}
To evaluate whether distributions differed between two independent groups, we used the Mann--Whitney U test, a nonparametric rank-based test that does not require assuming normality of the observations. The test assesses whether values from one group tend to be systematically larger or smaller than those from the other group by comparing sums of ranks after pooling the two groups (Figs.~S1C, S1D, 2B, 2C).

\paragraph{Permutation test.}
To compare community-level properties, such as species richness $N$, connectance $C$, and community complexity $NC$, among assembly conditions, we performed a permutation test on the mean difference. For each pairwise comparison, the test statistic was defined as the difference in the mean value of the focal metric between two groups:
\begin{linenomath*}
\[
T_{\mathrm{obs}} = \bar{X}_{A} - \bar{X}_{B},
\]
\end{linenomath*}
where $\bar{X}_{A}$ and $\bar{X}_{B}$ denote the mean of each property in compared conditions $A$ and $B$. To generate the null distribution expected in the absence of condition effects, we pooled data from two conditions and randomly permuted condition labels while keeping group sizes fixed. For each of $B$ random permutations, we recalculated the mean difference, denoted $T_{\mathrm{perm}}^{(k)}$ for the $k$-th permutation ($k=1,\ldots,B$). Let $b$ denote the number of permutations satisfying $|T_{\mathrm{perm}}^{(k)}| \ge |T_{\mathrm{obs}}|$. The two-sided $p$-value was computed with add-one smoothing as $p = (b+1)/(B+1)$. We used $B = 100{,}000$ permutations for each comparison (Figs.~4A and S4A).

\subsection{Quantifying community complexity and linear stability in the GLV model}

For each simulation replicate, species with abundances above a predefined persistence threshold at the final time point were regarded as coexisting. Then, we measured connectance $C$ as [20, 48]
\begin{linenomath*}
\[
C = \frac{L}{N(N-1)},
\]
\end{linenomath*}
where $N$ is the number of coexisting species and $L$ is the number of nonzero off-diagonal elements representing interspecific interactions in the interaction matrix $\alpha$. We adopted the product of diversity and connectance, $NC$, as a measure of community complexity [62, 63]. We also quantified a May-type complexity measure [20, 48], defined as
\begin{linenomath*}
\[
C_{\mathrm{May}} = \sigma \sqrt{NC},
\]
\end{linenomath*}
where $\sigma$ denotes the standard deviation of the nonzero off-diagonal elements of the Jacobian submatrix $J_{ij} = \partial f_i / \partial x_j$ of the GLV model (Eq.1). The element $J_{ij}$ describes how a small change in the abundance of species $j$ affects the growth rate of species $i$ locally around the focal state at time $T_{\mathrm{end}}$, providing a measure of linear asymptotic stability of community dynamics [20, 21, 22, 23, 24]. This index increases with the dimensionality of the coexisting community $N$, the density $C$ of interactions, and the variability $\sigma$ of interaction strength.

\section{Results}

\subsection{Skewed weak and Pareto-tailed strong interactions (SWAPS) distributions in mutualistic and predator--prey interactions}

We investigated whether a mutualistic network composed of angiosperm and pollinator species follows SW and heavy-tailed distributions (Fig.~1A bottom). Indeed, this network contains a large number of weak interactions and a few strong ones, as quantified by two independent approaches, visit-based and encounter-based (visit: Fig.~1B; encounter: Fig.~S1A). The skewness of the IS distribution was positive, indicating a significantly right-skewed deviation from the Gaussian distribution (visit: $Z_{\mathrm{skew}} = 114$, $p < 0.01$; encounter: $Z_{\mathrm{skew}} = 100$, $p < 0.01$; one-tailed test on the right), confirming SW. Next, to quantitatively evaluate the heavy tail, we estimated the Pareto exponent $\xi$ using the peak-over-threshold (POT) method with a generalized Pareto fit [58, 59, 60, 64] of the IS distribution above a threshold (Methods). A positive Pareto exponent $\xi$ in the generalized Pareto distribution indicates a heavier tail, hereafter referred to as Pareto-tailed strong interactions (PS), than the exponential distribution [64] (Fig.~1B, inset). As a result, the Pareto exponents $\xi$ in both approaches were significantly greater than zero and showed little dependence on a parameter used for the POT method (Fig.~S1B), indicating a heavy tail (Fig.~1C; Methods). These results quantitatively show that this mutualistic network follows the distribution of skewed weak and Pareto-tailed strong interactions (SWAPS).

To examine whether SWAPS distribution is common to multiple interaction types coexisting within a single community, we analyzed an agroecosystem containing both mutualistic and trophic interactions (Fig.~1D). For mutualistic IS, skewness was positive, indicating a significantly right-skewed deviation from the Gaussian distribution and showing SW ($Z_{\mathrm{skew}} = 74.4$, $p < 0.01$, one-tailed test to the right) (Fig.~1E left). Moreover, the POT-based estimate of $\xi$ was positive, supporting a heavy tail, although the lower bound of the 95\% bootstrap confidence interval was non-positive (Fig.~1C; Methods). For predator--prey interactions, the overall IS distribution was similarly right-skewed (Fig.~1E right). While predator IS exhibited a bimodal distribution according to two taxonomic groups, invertebrates (e.g., aphids and leaf miners) and vertebrates (e.g., birds and rodents), the IS of each predator group exhibited a unimodal distribution on a logarithmic scale (Fig.~1E right, deep and light green, respectively), indicating a significantly right-skewed deviation from the Gaussian distribution (invertebrates: $Z_{\mathrm{skew}} = 90.7$, $p < 0.01$; vertebrates: $Z_{\mathrm{skew}} = 156$, $p < 0.01$; one-tailed test to the right for skewness). Additionally, the estimated Pareto exponent $\xi$ was significantly positive in each unimodal distribution, showing PS (Fig.~1C). These results statistically reveal SWAPS distribution across mutualistic and trophic interactions within a single community, suggesting a common principle that generates SWAPS distribution regardless of interaction type.

\subsection{Taxonomic specificity and conservatism on interaction strength}

The interaction strength distribution differed markedly with taxonomic combinations in the agroecosystem (Fig.~2A). For example, among taxonomic pairs with abundant mutualistic interactions ($>10$), IS distributions were substantially higher between Asteraceae species and Syrphidae pollinators, lower between the same plant family and Nymphalidae pollinators, and intermediate between Apiaceae species and Syrphidae pollinators. To quantitatively assess the taxonomic specificity of IS, we calculated IS averaged for each taxonomic pair by averaging IS values within each combination. We found that IS differed significantly in approximately half of the top 10 most frequent taxonomic combinations (Fig.~2B; 22 comparisons out of $\binom{10}{2} = 45$ pairwise comparisons with adjusted $p$-values $< 0.05$; Mann--Whitney U test with multiple-comparisons correction), indicating that interaction strength is substantially specific to the interacting taxa.

Moreover, the strength of mutualistic interaction exhibited smaller variation within each taxonomic pair (Fig.~2A, e.g., Apiaceae--Syrphidae, Asteraceae--Calliphoridae, Fabaceae--Apidae) than in the whole ecosystem (Fig.~2A, top). Thus, we quantitatively examined whether this small variation indicates taxonomic conservatism, which remains uninvestigated for IS. We calculated the coefficient of variation (CV) for IS values within each pair of taxonomic combinations. The CV values were then compared to those derived from permuted communities, in which IS values were randomly shuffled while maintaining the original interaction topology. For both mutualistic and trophic interactions, the CVs for taxonomic group combinations were significantly smaller than those in shuffled communities (Mann--Whitney U test, $p < 0.01$; Figs.~2C, S1D). In the plant--pollinator network, the CVs tended to be consistently smaller than the randomized case, with no significant difference (Mann--Whitney U test, $p = 0.42$; Fig.~S1C). Therefore, IS is widely distributed depending on the interacting taxa, but varies little within each taxonomic combination. These results support the taxonomic specificity and conservatism of IS.

Taxonomic specificity and conservatism prompted us to examine whether IS averaged for each taxonomic combination also indicates wide variation contributing to SWAPS distribution formation. Strikingly, IS averaged at both the family and order levels in the agroecosystem and plant--pollinator network (Fig.~2D) significantly showed deviations in skewness from the Gaussian distribution ($p < 0.01$, one-tailed test to the right, respectively). Moreover, this family-level IS indicated positive Pareto exponent $\xi$ in the plant--pollinator network and trophic interaction of the agroecosystem, while $\xi$ at the family level was non-positive in the mutualistic interaction of the agroecosystem (Fig.~S1E). Overall, SWAPS distribution consistently emerges in interaction strength at the taxonomic and species levels of the observed data, except for heavy-tailedness at the taxonomic level in the mutualistic interaction of the agroecosystem.

\subsection{Generative principles of SWAPS distribution emergence}

We then investigated whether the SWAPS distribution observed in mutualistic and trophic interactions (Figs.~1B, E) can emerge in community assembly simulations based on the GLV model. We incorporated repeated immigration of taxonomically close species possessing the observed taxonomic conservatism in interaction strength (Fig.~2C) and distant species lacking this trait into each assembling community. The assembling communities included multiple interaction types, including mutualism, predator--prey/parasitism, commensalism, interference competition, and amensalism (Fig.~3A). As generations of assembly simulations progressed, coexisting species richness continuously increased and the clade-size distribution became uneven, exhibiting a few large clades and many small ones (Figs.~S2A left and S2D), similarly to the observed communities (Fig.~S2A right). The assembled communities showed that the skewness of the IS distribution for each interaction type significantly deviated in the positive direction from the Gaussian distribution, indicating SW distribution ($p < 0.01$, one-tailed test to the right), in almost all evolutionary simulations (Fig.~3B left, Table~1). Furthermore, across fifty independently assembled communities, the median estimated Pareto exponent $\xi$ was positive for mutualism, predator--prey, and amensalism (Fig.~3B right), whereas the lower bound of its bootstrap 95\% confidence interval for each community was not always positive (Fig.~3B right, bottom). These results indicate that this community assembly process with multiple interaction types and taxonomic conservatism consistently generates SWAPS distribution in mutualism, predator--prey, and amensalism, alongside multispecies coexistence.

We next examined whether SWAPS distribution identified at the taxonomic-group level in the data analysis (Fig.~2D) was also observed at the lineage level corresponding to taxonomically close species in the simulations (Fig.~S2B). The tendency for IS distribution to be significantly skewed toward weak interactions was largely preserved after lineage-level averaging, except for competition ($p < 0.05$, one-sided test; Fig.~S2C left). Furthermore, communities with positive $\xi$ emerged more frequently than those with negative $\xi$, except for competition and commensalism, although there were not many communities capable of quantifying $\xi$ (Fig.~S2C right). Together, these results indicate SWAPS distribution emergence at the lineage level as well as the species level.

To understand which interaction types facilitate SWAPS distribution emergence, we first simulated community assembly with trophic interactions alone, which were observed in this study (Fig.~1E) and have been referred to in observations of SWAPS distribution in food webs [18, 19]. Under immigration of both distant and close species, SW always emerged through community assembly (Figs.~S3A, B left). However, fewer replicates showed positive $\xi$ for the median and the lower bound of its bootstrap 95\% confidence interval than under multiple-interaction-type assembly (Figs.~S3A, B right). Next, we attempted community assembly with mutualistic and trophic interactions, which we analyzed in datasets (Fig.~1E). These assemblies led to low reproducibility of positive skewness (less than 50\% in Table~1) and saturation of low species richness and few interactions in each community (Fig.~S2D), preventing us from estimating $\xi$ (Fig.~S3D). Therefore, SW emerges primarily under trophic-alone or multiple-interaction-type settings, while heavy-tail emergence requires not only trophic interactions but also the other four interaction types to improve reproducibility.

To further examine the role of taxonomic conservatism in SWAPS distribution emergence, we introduced immigration of distant species alone without close species in three community assembly settings: (i) trophic interactions alone, (ii) mutualistic and trophic interactions, and (iii) all five interaction types. We found that the IS distribution severely lacked positive skewness in any setting (28\% at most; Table~1, Figs.~S3F, S3H), compared with simulations in the presence of taxonomic conservatism. Regarding the heavy tail, under trophic interaction alone, all replicates showed negative $\xi$ (Table~1, Figs.~S3E--F). In assemblies involving mutualistic and trophic interactions or multiple interaction types, too few interactions were established within each interaction type for tail estimation, preventing this distribution analysis (Figs.~S3G--H). These results indicate the generative principles of SWAPS distribution emergence: taxonomic conservatism promotes the emergence of SW distribution (Fig.~3B, Table~1), whereas multiple interaction types as well as taxonomic conservatism jointly support the emergence of a heavy tail (Figs.~3B and S3I--J) through the assembly of complex communities.

\subsection{SWAPS distribution co-emerges with community complexity}

We finally asked how these assembly conditions generating SWAPS distribution affect community diversity and complexity. To this end, we evaluated diversity by the number of coexisting species $N$ and the Shannon--Wiener index [65], and complexity by the product of $N$ and connectance $C$ (Methods). We found that mean connectance $C$ was significantly at least 6.73 times higher under taxonomic conservatism, given SWAPS distribution with multiple interaction types or predator--prey alone (Table~1), than under its absence (Fig.~4A center, two-sided permutation tests with Holm correction). In contrast, the effects of taxonomic conservatism on community diversity $N$ and the Shannon--Wiener index were weak, with the mean differing by no more than 1.26 times and without consistent directional effects across assembly conditions (Figs.~4A left, S4A left; two-sided permutation tests with Holm correction). By multiplying these effects, community complexity $NC$ was at least 5.40 times higher under taxonomic conservatism than in its absence (Fig.~4A right, two-sided permutation tests with Holm correction). Additionally, the classical community complexity $\sigma \sqrt{NC}$ [10, 48], where $\sigma$ denotes the standard deviation of the nonzero Jacobian elements of the GLV model, exhibited the same tendency (Fig.~S4A center and right; Methods). Given taxonomic conservatism, we further found that assemblies under predator--prey and mutualistic interactions showed significantly lower diversity $N$ (below 100 species) and community complexity $NC$ than the other conditions, predator--prey alone and multiple interaction types (Figs.~4A, S4B). Together, these results suggest that interaction type composition and taxonomic conservatism are associated primarily with diversity and connectance, respectively, thereby jointly contributing to higher community complexity.

For diverse communities, i.e., those above 100 species assembled under multiple interaction types or predator--prey alone (Fig.~4A), we examined the relationship between complexity and SWAPS distribution emergence, for which taxonomic conservatism is indispensable (Figs.~3B and S3I--J). First, we found that $NC$ discontinuously increases above the threshold of the SW indicator $Z_{\mathrm{skew}}$ (dashed line, Fig.~4B left). Second, in communities with higher $NC$ (Fig.~4B, $NC \ge 10$; Fig.~4C, $[10,\infty)$), the heavy-tail indicator $\xi$ was variable (Fig.~4B right) but tended to be positive (deep red, Figs.~4C and S4B) more frequently than negative (pale red, Figs.~4C and S4B). Moreover, we found that the relative frequency of communities with heavy-tailed IS distributions, i.e., positive $\xi$, to those without heavy tails, i.e., negative $\xi$, became greater as $NC$ increased (Fig.~4C, predator--prey alone). This trend was also observed in May's complexity $\sigma \sqrt{NC}$ (Fig.~S4C, except for competition). Taken together, these results suggest that complex communities are always SW and tend to be heavy-tailed.

\section{Discussion}

In summary, we quantitatively showed that interaction strength in both predator--prey and mutualistic relationships is distributed with both skewed weak and Pareto-tailed strong interactions (SWAPS) through analyses of two empirical communities (Fig.~1). Interaction strength indicates taxonomic conservatism (Fig.~2), which constitutes the generative principle of SWAPS distribution with multiple interaction types through community assembly simulations (Fig.~3). The emergence of SWAPS distribution was correlated with increasing community complexity (Fig.~4).

In a mutualistic community and a mixed community including mutualistic and predator--prey interactions, it has remained unclear whether interaction strength distributions are heavy-tailed and skewed toward weak interactions due to a lack of quantitative analysis that distinguishes these two distributional features. Our quantitative analyses combining the skewness indicator $Z_{\mathrm{skew}}$ with the tail-shape parameter $\xi$ revealed that both trophic and mutualistic interactions were significantly skewed toward weak interactions and heavy-tailed (Fig.~1C). Moreover, averaging IS at the level of taxa revealed taxon specificity and conservatism (Figs.~2B, C); IS differed markedly among taxonomic combinations, yet varied less within each combination, indicating that interaction strengths are not assigned arbitrarily across ecological networks but are at least partly constrained by the identities of the interacting taxa. Our taxonomic-pair-based analysis can be applied with relatively limited data compared with approaches based on branch lengths in phylogenetic trees [31, 66], making it a practical complement for re-examining existing ecological datasets. The presented pipeline quantifying SWAPS distribution at species and taxonomic levels is widely applicable to interaction strengths inferred from recent time-series data, such as those estimated from environmental DNA in fish communities [27, 28] and mesocosm communities [67], enabling tests of whether SWAPS distribution is shared across different taxonomic kingdoms and ecological contexts.

While the underlying generative principle of SWAPS distribution has remained unclear, our community assembly simulations of the GLV model show that not only taxonomic conservatism but also multiple interaction types, not limited to predator--prey (trophic) and mutualistic interactions, jointly drive SWAPS distribution emergence (Fig.~3A). This generative principle will likely be examined extensively in the future through research into which combinations or relative frequencies of interaction types are sufficient for SWAPS distribution emergence, as well as through other frameworks, such as agent-based models and models with explicit spatial structure [68, 69], thereby verifying robustness. We should note that interaction networks generated in the present assembly simulations did not fully account for the topological properties observed in empirical networks, possibly due to the lack of mechanisms that amplify degree heterogeneity during lineage diversification. Future models should therefore incorporate additional biological constraints on link acquisition or degree structure, such as phylogenetic conservatism in partner range, trait-based partner matching, or mechanisms that generate and maintain generalist--specialist differences. This extension would allow tests of whether the same assembly framework can reproduce both SWAPS distribution and empirical topological organization.

Importantly, SWAPS distribution emergence at the species level is associated with increasing community complexity (Figs.~4B--C and S4A), whereas earlier theoretical studies often assumed ISs or Jacobians follow Gaussian distributions that lack SW and heavy tails [20, 48]. Moreover, SWAPS distribution at the lineage level simultaneously emerges (Fig.~3C). These associations provide a future avenue to examine the causality between SWAPS distribution and community functions during community assembly by extensively tracing how perturbations on the generative principles of SWAPS distribution (Figs.~3 and S3) influence community diversity and complexity via changes in skewness and heavy tails. Further analyses of such associations throughout various community data and models would clarify whether SWAPS distribution at species and lineage levels provides new perspectives for understanding diversity and complexity. Given that skewness toward weak interactions has often been attributed to community stability [20, 21, 22, 23, 24], future research combining the presented quantitative analyses may further find associations of SWAPS distribution with other functions of complex communities, including stability and resilience.

\section*{Acknowledgement}

We thank Katsuyoshi Matsushita (Hiroshima University, Japan), Katsuhiko Yoshida (National Institute for Environmental Studies, Japan), and members of the Complexity Life Mathematics laboratory (Hiroshima University, Japan) for valuable discussions and suggestions. This work is supported by the Japan Science and Technology Agency (SPRING, Grant Number JPMJSP2108 to T. H. and CREST JPMJCR2121 to K. F.) and the Japan Society for the Promotion of Science (26H00457 to K. F.).

\section*{Author Contributions}

Conceptualization: T.H. and K.F.; Investigation: T.H. and T.A.; Methodology: T.H.; Data Curation: T.H.; Writing -- original draft: T.H. and K.F.; Writing -- review \& editing: T.H., T.A., and K.F.

\section*{Declaration of Interests}

The authors declare no competing interests.

\section*{References}
\begin{enumerate}
\item Ricklefs, R. E. (2004). A comprehensive framework for global patterns in biodiversity. \textit{Ecology Letters}, 7(1), 1--15. doi: \url{https://doi.org/10.1046/j.1461-0248.2003.00554.x}
\item Levin, S. A. (1998). Ecosystems and the biosphere as complex adaptive systems. \textit{Ecosystems}, 1, 431--436. doi: \url{https://doi.org/10.1007/s100219900037}
\item Dunne, J. A., Williams, R. J., \& Martinez, N. D. (2002). Food-web structure and network theory: The role of connectance and size. \textit{Proceedings of the National Academy of Sciences of the United States of America}, 99(20), 12917--12922. doi: \url{https://doi.org/10.1073/pnas.192407699}
\item Olesen, J. M., \& Jordano, P. (2002). Geographic patterns in plant--pollinator mutualistic networks. \textit{Ecology}, 83(9), 2416--2424. doi: \url{https://doi.org/10.2307/3071803}
\item Rezende, E. L., Jordano, P., \& Bascompte, J. (2007). Effects of phenotypic complementarity and phylogeny on the nested structure of mutualistic networks. \textit{Oikos}, 116(11), 1919--1929. doi: \url{https://doi.org/10.1111/j.0030-1299.2007.16029.x}
\item Holland, J. N., \& DeAngelis, D. L. (2009). Consumer-resource theory predicts dynamic transitions between outcomes of interspecific interactions. \textit{Ecology Letters}, 12(12), 1357--1366. doi: \url{https://doi.org/10.1111/j.1461-0248.2009.01390.x}
\item Bascompte, J. (2010). Structure and dynamics of ecological networks. \textit{Science}, 329(5993), 765--766. doi: \url{https://doi.org/10.1126/science.1194255}
\item Hutchinson, G. E. (1959). Homage to Santa Rosalia, or why are there so many kinds of animals? \textit{The American Naturalist}, 93(870), 145--159. doi: \url{https://doi.org/10.1086/282070}
\item Calcagno, V., Jarne, P., Loreau, M., Mouquet, N., \& David, P. (2017). Diversity spurs diversification in ecological communities. \textit{Nature Communications}, 8, 15810. doi: \url{https://doi.org/10.1038/ncomms15810}
\item Landi, P., Minoarivelo, H. O., Br\"annstr\"om, \AA., Hui, C., \& Dieckmann, U. (2018). Complexity and stability of ecological networks: A review of the theory. \textit{Population Ecology}, 60(4), 319--345. doi: \url{https://doi.org/10.1007/s10144-018-0628-3}
\item Olesen, J. M., Bascompte, J., Dupont, Y. L., \& Jordano, P. (2007). The modularity of pollination networks. \textit{Proceedings of the National Academy of Sciences of the United States of America}, 104(50), 19891--19896. doi: \url{https://doi.org/10.1073/pnas.0706375104}
\item Xu, Y., Jord\'an, F., Zhou, M., Huo, X., Cai, Y., Ur Rehman, S., \& Sun, J. (2024). Global variability of degree distribution in marine food webs. \textit{Diversity and Distributions}, 30, e13927. doi: \url{https://doi.org/10.1111/ddi.13927}
\item Laska, M. S., \& Wootton, J. T. (1998). Theoretical concepts and empirical approaches to measuring interaction strength. \textit{Ecology}, 79(2), 461--476. doi: \url{https://doi.org/10.1890/0012-9658(1998)079[0461:TCAEAT]2.0.CO;2}
\item Berlow, E. L., Neutel, A.-M., Cohen, J. E., de Ruiter, P. C., Ebenman, B., Emmerson, M., Fox, J. W., Jansen, V. A. A., Jones, J. I., Kokkoris, G. D., Logofet, D. O., McKane, A. J., Montoya, J. M., \& Petchey, O. (2004). Interaction strengths in food webs: Issues and opportunities. \textit{Journal of Animal Ecology}, 73(3), 585--598. doi: \url{https://doi.org/10.1111/j.0021-8790.2004.00833.x}
\item Berlow, E. L., Navarrete, S. A., Briggs, C. J., Power, M. E., \& Menge, B. A. (1999). Quantifying variation in the strengths of species interactions. \textit{Ecology}, 80(7), 2206--2224. doi: \url{https://doi.org/10.1890/0012-9658(1999)080[2206:QVITSO]2.0.CO;2}
\item Bascompte, J., Meli\'an, C. J., \& Sala, E. (2005). Interaction strength combinations and the overfishing of a marine food web. \textit{Proceedings of the National Academy of Sciences of the United States of America}, 102(15), 5443--5447. doi: \url{https://doi.org/10.1073/pnas.0501562102}
\item Wootton, J. T., \& Emmerson, M. (2005). Measurement of interaction strength in nature. \textit{Annual Review of Ecology, Evolution, and Systematics}, 36, 419--444. doi: \url{https://doi.org/10.1146/annurev.ecolsys.36.091704.175535}
\item Sala, E., \& Graham, M. H. (2002). Community-wide distribution of predator--prey interaction strength in kelp forests. \textit{Proceedings of the National Academy of Sciences of the United States of America}, 99(6), 3678--3683. doi: \url{https://doi.org/10.1073/pnas.052028499}
\item Emmerson, M. C., \& Raffaelli, D. (2004). Predator--prey body size, interaction strength and the stability of a real food web. \textit{Journal of Animal Ecology}, 73(3), 399--409. doi: \url{https://doi.org/10.1111/j.0021-8790.2004.00818.x}
\item Allesina, S., \& Tang, S. (2012). Stability criteria for complex ecosystems. \textit{Nature}, 483(7388), 205--208. doi: \url{https://doi.org/10.1038/nature10832}
\item van Altena, C., Hemerik, L., \& de Ruiter, P. C. (2016). Food web stability and weighted connectance: The complexity--stability debate revisited. \textit{Theoretical Ecology}, 9(1), 49--58. doi: \url{https://doi.org/10.1007/s12080-015-0291-7}
\item Mittelbach, G. G. (2023). \textit{Community Ecology} (2nd ed.; H. Kadowaki, M. Yamamichi, \& S. Utsumi, Trans.). Maruzen Publishing.
\item Rooney, N., \& McCann, K. S. (2012). Integrating food web diversity, structure and stability. \textit{Trends in Ecology \& Evolution}, 27(1), 40--46. doi: \url{https://doi.org/10.1016/j.tree.2011.09.001}
\item Emmerson, M., \& Yearsley, J. M. (2004). Weak interactions, omnivory and emergent food-web properties. \textit{Proceedings of the Royal Society B: Biological Sciences}, 271(1537), 397--405. doi: \url{https://doi.org/10.1098/rspb.2003.2592}
\item Bascompte, J., Jordano, P., \& Olesen, J. M. (2006). Asymmetric coevolutionary networks facilitate biodiversity maintenance. \textit{Science}, 312(5772), 431--433. doi: \url{https://doi.org/10.1126/science.1123412}
\item V\'azquez, D. P., Lom\'ascolo, S. B., Maldonado, M. B., Chacoff, N. P., Dorado, J., Stevani, E. L., \& Vitale, N. L. (2012). The strength of plant--pollinator interactions. \textit{Ecology}, 93(4), 719--725. doi: \url{https://doi.org/10.1890/11-1356.1}
\item Ushio, M., Hsieh, C.-H., Masuda, R., Deyle, E. R., Ye, H., Chang, C.-W., Sugihara, G., \& Kondoh, M. (2018). Fluctuating interaction network and time-varying stability of a natural fish community. \textit{Nature}, 554, 360--363. doi: \url{https://doi.org/10.1038/nature25504}
\item Ushio, M., Sado, T., Fukuchi, T., Sasano, S., Masuda, R., Osada, Y., \& Miya, M. (2023). Temperature sensitivity of the interspecific interaction strength of coastal marine fish communities. \textit{eLife}, 12, RP85795. doi: \url{https://doi.org/10.7554/eLife.85795.3}
\item G\'omez, J. M., Verd\'u, M., \& Perfectti, F. (2010). Ecological interactions are evolutionarily conserved across the entire tree of life. \textit{Nature}, 465(7300), 918--921. doi: \url{https://doi.org/10.1038/nature09113}
\item Aizen, M. A., Gleiser, G., Sabatino, M., Gilarranz, L. J., Bascompte, J., \& Verd\'u, M. (2016). The phylogenetic structure of plant--pollinator networks increases with habitat size and isolation. \textit{Ecology Letters}, 19(1), 29--36. doi: \url{https://doi.org/10.1111/ele.12539}
\item Rafferty, N. E., \& Ives, A. R. (2013). Phylogenetic trait-based analyses of ecological networks. \textit{Ecology}, 94(10), 2321--2333. doi: \url{https://doi.org/10.1890/12-1948.1}
\item Fritschie, K. J., Cardinale, B. J., Alexandrou, M. A., \& Oakley, T. H. (2014). Evolutionary history and the strength of species interactions: Testing the phylogenetic limiting similarity hypothesis. \textit{Ecology}, 95(5), 1407--1417. doi: \url{https://doi.org/10.1890/13-0986.1}
\item de Ruiter, P. C., Neutel, A.-M., \& Moore, J. C. (1995). Energetics, patterns of interaction strengths, and stability in real ecosystems. \textit{Science}, 269(5228), 1257--1260. doi: \url{https://doi.org/10.1126/science.269.5228.1257}
\item Pimm, S. L. (1980). Food web design and the effect of species deletion. \textit{Oikos}, 35(2), 139--149. doi: \url{https://doi.org/10.2307/3544422}
\item V\'azquez, D. P., Morris, W. F., \& Jordano, P. (2005). Interaction frequency as a surrogate for the total effect of animal mutualists on plants. \textit{Ecology Letters}, 8(10), 1088--1094. doi: \url{https://doi.org/10.1111/j.1461-0248.2005.00810.x}
\item Bimler, M. D., Mayfield, M. M., Martyn, T. E., \& Stouffer, D. B. (2023). Estimating interaction strengths for diverse horizontal systems using performance data. \textit{Methods in Ecology and Evolution}, 14(4), 1075--1089. doi: \url{https://doi.org/10.1111/2041-210X.14068}
\item Novak, M., Wootton, J. T., Doak, D. F., Emmerson, M., Estes, J. A., \& Tinker, M. T. (2016). Characterizing species interactions to understand press perturbations: What is the community matrix? \textit{Annual Review of Ecology, Evolution, and Systematics}, 47, 409--432. doi: \url{https://doi.org/10.1146/annurev-ecolsys-032416-010215}
\item Robinson, J. V., \& Valentine, W. D. (1979). The concepts of elasticity and resilience in ecological systems. \textit{Ecology}, 60(6), 1312--1314.
\item Yodzis, P. (1988). The indeterminacy of ecological interactions as perceived through perturbation experiments. \textit{Ecology}, 69(2), 508--515. doi: \url{https://doi.org/10.2307/1940449}
\item Drake, J. A. (1990). The mechanics of community assembly and succession. \textit{Journal of Theoretical Biology}, 147(2), 213--233. doi: \url{https://doi.org/10.1016/S0022-5193(05)80053-0}
\item Poley, L., Baron, J. W., \& Galla, T. (2023). Generalized Lotka--Volterra model with hierarchical interactions. \textit{Physical Review E}, 107, 024313. doi: \url{https://doi.org/10.1103/PhysRevE.107.024313}
\item Ogushi, T., Kondo, M., \& Namba, T. (Eds.). (2009). \textit{Community Ecology 3: Deciphering Biological Interaction Networks}. Kyoto University Press.
\item Chesson, P., \& Huntly, N. (1997). The roles of harsh and fluctuating conditions in the dynamics of ecological communities. \textit{The American Naturalist}, 150(5), 519--553. doi: \url{https://doi.org/10.1086/286080}
\item Adler, P. B., HilleRisLambers, J., \& Levine, J. M. (2007). A niche for neutrality. \textit{Ecology Letters}, 10(2), 95--104. doi: \url{https://doi.org/10.1111/j.1461-0248.2006.00996.x}
\item Yoshida, K., Hata, K., Kawakami, K., et al. (2023). Predicting ecosystem changes by a new model of ecosystem evolution. \textit{Scientific Reports}, 13, 15353. doi: \url{https://doi.org/10.1038/s41598-023-42529-9}
\item Drossel, B., Higgs, P. G., \& McKane, A. J. (2001). The influence of predator--prey population dynamics on the long-term evolution of food web structure. \textit{Journal of Theoretical Biology}, 208(1), 91--107. doi: \url{https://doi.org/10.1006/jtbi.2000.2203}
\item Tokita, K., \& Yasutomi, A. (2003). Emergence of a complex and stable network in a model ecosystem with extinction and mutation. \textit{Theoretical Population Biology}, 63(2), 131--146. doi: \url{https://doi.org/10.1016/S0040-5809(02)00038-2}
\item May, R. M. (1972). Will a large complex system be stable? \textit{Nature}, 238(5364), 413--414. doi: \url{https://doi.org/10.1038/238413a0}
\item Novella-Fernandez, R., Rodrigo, A., Arnan, X., \& Bosch, J. (2019). Interaction strength in plant--pollinator networks: Are we using the right measure? \textit{PLOS ONE}, 14(12), e0225930. doi: \url{https://doi.org/10.1371/journal.pone.0225930}
\item Pocock, M. J. O., Evans, D. M., \& Memmott, J. (2012). The robustness and restoration of a network of ecological networks. \textit{Science}, 335(6071), 973--977. doi: \url{https://doi.org/10.1126/science.1214915}
\item Hofbauer, J., \& Sigmund, K. (1998). \textit{Evolutionary Games and Population Dynamics}. Cambridge University Press.
\item Seno, H. (2007). \textit{Mathematical Biology: Introduction to Mathematical Modeling of Population Dynamics}. Kyoritsu Shuppan.
\item Cahill, J. F., Jr., Kembel, S. W., Lamb, E. G., \& Keddy, P. A. (2008). Does phylogenetic relatedness influence the strength of competition among vascular plants? \textit{Perspectives in Plant Ecology, Evolution and Systematics}, 10(1), 41--50. doi: \url{https://doi.org/10.1016/j.ppees.2007.10.001}
\item Barab\'asi, A.-L., \& Albert, R. (1999). Emergence of scaling in random networks. \textit{Science}, 286(5439), 509--512. doi: \url{https://doi.org/10.1126/science.286.5439.509}
\item Adler, P. B., Ellner, S. P., \& Levine, J. M. (2010). Coexistence of perennial plants: An embarrassment of niches. \textit{Ecology Letters}, 13(8), 1019--1029. doi: \url{https://doi.org/10.1111/j.1461-0248.2010.01496.x}
\item Comita, L. S., Muller-Landau, H. C., Aguilar, S., \& Hubbell, S. P. (2010). Asymmetric density dependence shapes species abundances in a tropical tree community. \textit{Science}, 329(5989), 330--332. doi: \url{https://doi.org/10.1126/science.1190772}
\item Jarque, C. M., \& Bera, A. K. (1987). A test for normality of observations and regression residuals. \textit{International Statistical Review}, 55(2), 163--172. doi: \url{https://doi.org/10.2307/1403192}
\item Davison, A. C., \& Smith, R. L. (1990). Models for exceedances over high thresholds. \textit{Journal of the Royal Statistical Society: Series B (Methodological)}, 52(3), 393--442. doi: \url{https://doi.org/10.1111/j.2517-6161.1990.tb01796.x}
\item Pickands, J., III. (1975). Statistical inference using extreme order statistics. \textit{The Annals of Statistics}, 3(1), 119--131. doi: \url{https://doi.org/10.1214/aos/1176343003}
\item Balkema, A. A., \& de Haan, L. (1974). Residual life time at great age. \textit{The Annals of Probability}, 2(5), 792--804. doi: \url{https://doi.org/10.1214/aop/1176996548}
\item Dupuis, D. J. (1999). Exceedances over high thresholds: A guide to threshold selection. \textit{Extremes}, 1(3), 251--261. doi: \url{https://doi.org/10.1023/A:1009914915709}
\item Keitt, T. H. (1997). Stability and complexity on a lattice: Coexistence of species in an individual-based food web model. \textit{Ecological Modelling}, 102(2--3), 243--258. doi: \url{https://doi.org/10.1016/S0304-3800(97)00059-8}
\item Jacquet, C., Moritz, C., Morissette, L., et al. (2016). No complexity--stability relationship in empirical ecosystems. \textit{Nature Communications}, 7, 12573. doi: \url{https://doi.org/10.1038/ncomms12573}
\item Rostami, M., Adam, M. B., Yahya, M. H., \& Ibrahim, N. A. (2018). Slice sampler algorithm for generalized Pareto distribution. \textit{Hacettepe Journal of Mathematics and Statistics}, 47(4), 1--32. doi: \url{https://doi.org/10.15672/HJMS.2017.441}
\item Margalef, R. (1957). La teor\'{i}a de la informaci\'{o}n en ecolog\'{i}a. \textit{Memorias de la Real Academia de Ciencias y Artes de Barcelona}, 32(13), 373--449.
\item Peralta, G. (2016). Merging evolutionary history into species interaction networks. \textit{Functional Ecology}, 30(12), 1917--1925. doi: \url{https://doi.org/10.1111/1365-2435.12669}
\item Hashimoto, K., Hayasaka, D., Eguchi, Y., et al. (2024). Multifaceted effects of variable biotic interactions on population stability in complex interaction webs. \textit{Communications Biology}, 7, 1309. doi: \url{https://doi.org/10.1038/s42003-024-06948-2}
\item Zelnik, Y. R., Solomon, S., \& Yaari, G. (2015). Species survival emerges from rare events of individual migration. \textit{Scientific Reports}, 5, 7877. doi: \url{https://doi.org/10.1038/srep07877}
\item Dobramysl, U., Mobilia, M., Pleimling, M., \& T\"auber, U. C. (2018). Stochastic population dynamics in spatially extended predator--prey systems. \textit{Journal of Physics A: Mathematical and Theoretical}, 51(6), 063001. doi: \url{https://doi.org/10.1088/1751-8121/aa95c7}
\end{enumerate}

\begin{figure}[p]
\centering
\includegraphics[width=\textwidth,height=0.76\textheight,keepaspectratio]{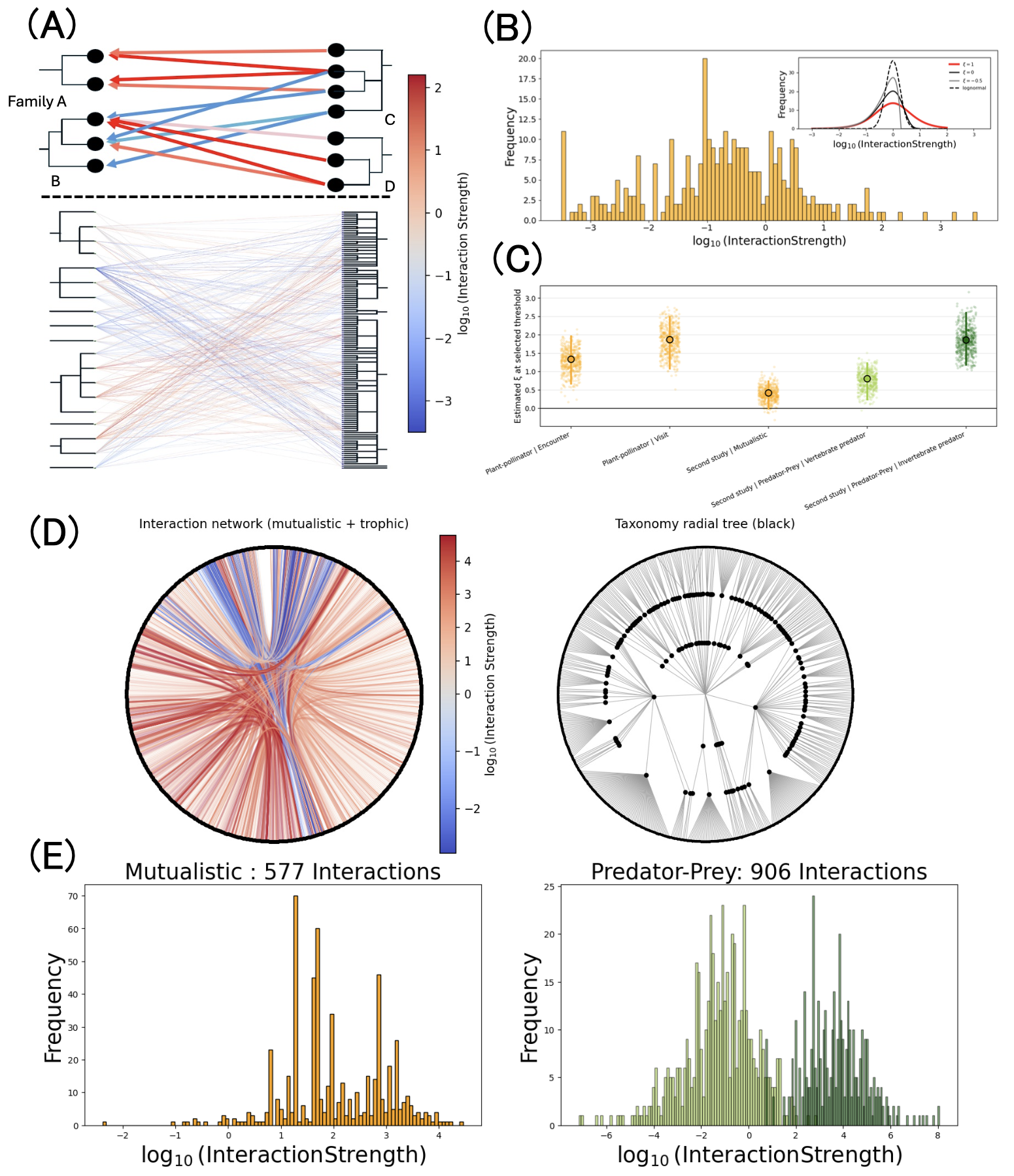}
\caption{\textbf{SWAPS interspecies interactions in empirical data.} Plant--pollinator mutualistic interaction network (A) and the frequency distributions of interaction strength (IS) (B; quantified by visit-based approaches). (C) The tail shape parameter $\xi$ of IS distribution quantified by different sampling methods in plant--pollinator network (the two on the left) and for different interaction types and taxa in agroecosystem (the three on the right), based on peak-over-threshold (POT) theorem. The median (black outlined circle) across replicates with bootstrap 95\% confidence intervals (vertical line) and the replicate-level estimates (points) are shown. Agroecosystem network (D, left), the taxonomic tree of the analyzed species (D, right), and the frequency distribution of IS (E; mutualistic [left panel], predator--prey [right]). The position of each species (black point) on the circle at the left panel in D is identical with that on the outermost circle at the right panel, while the higher taxa (families and orders, shown by points) are positioned on the inner circles. Predators were separated into vertebrates (light green) and invertebrates (deep green) in E. IS is shown by color (inset in A and D, left). The x-axis is log-transformed (B and E).}
\label{fig:fig1}
\end{figure}

\begin{figure}[p]
\centering
\includegraphics[width=\textwidth,height=0.78\textheight,keepaspectratio]{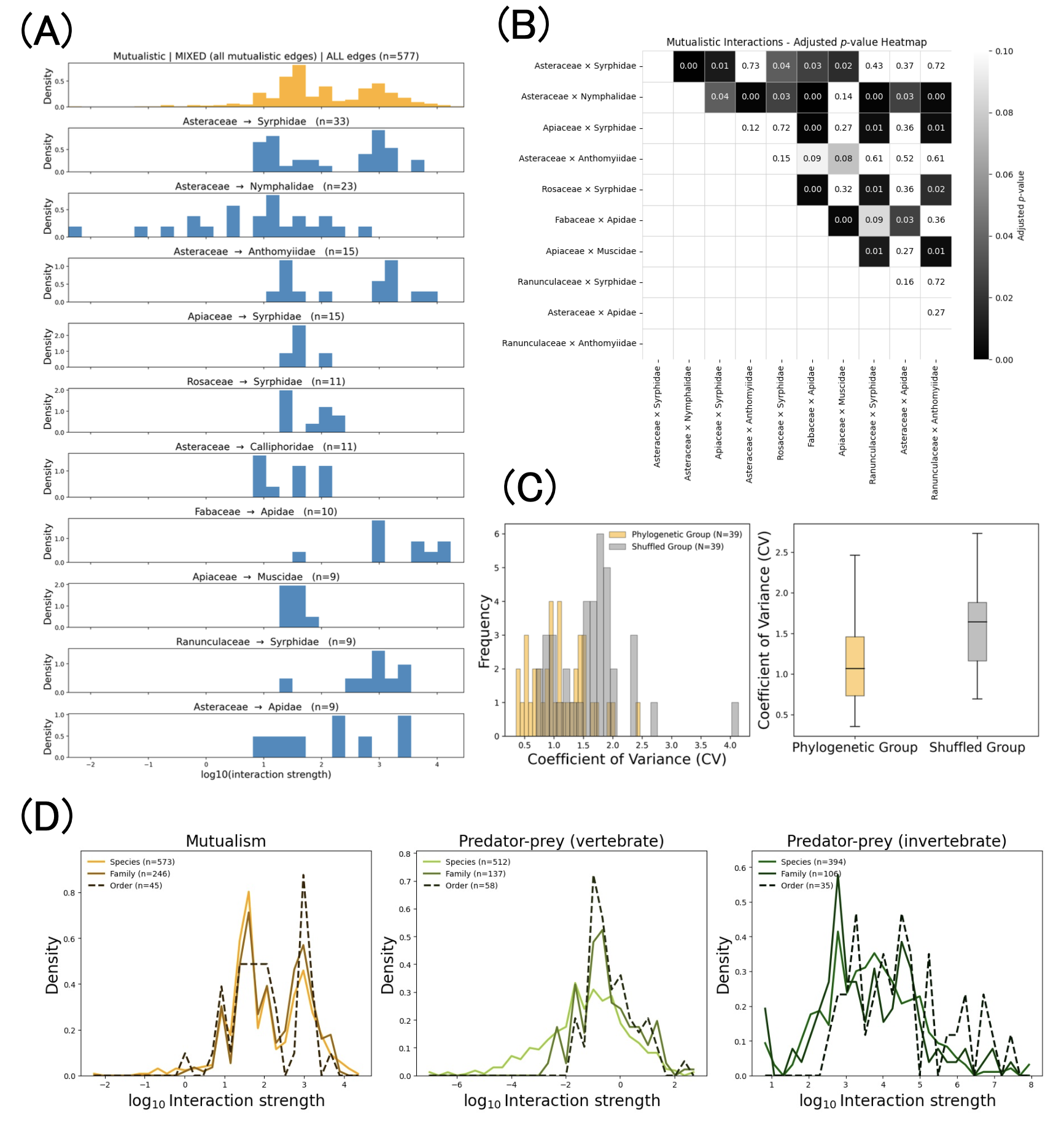}
\caption{\textbf{Taxonomic specificity and conservatism on interaction strength.} (A) Probability density distribution of mutualistic IS in the whole agroecosystem (top panel: yellow) and within each of the top 10 most frequent combinations of families (remaining panels: blue) with more than 9 interaction pairs. (B) Heatmap of $p$-values (adjusted for multiple comparisons) from Mann--Whitney U tests comparing IS averaged within these top 10 combinations of families. (C) Coefficient of variation (CV) of the averaged IS within the families with four or more mutualistic IS (orange) and shuffled groups (grey), shown as a histogram (left) and box plot (right). For the shuffled groups, IS values were randomly permuted among all links while keeping the network topology. (D) Probability density distributions of IS at the three taxonomic levels (species, family and order indicated by line style) of mutualism (left panel), predator--prey with vertebrate predator (center panel), and predator--prey with invertebrate predator (right panel). The family- and order-level IS were averaged within the respective taxonomic combinations.}
\label{fig:fig2}
\end{figure}

\begin{figure}[p]
\centering
\includegraphics[width=\textwidth,height=0.74\textheight,keepaspectratio]{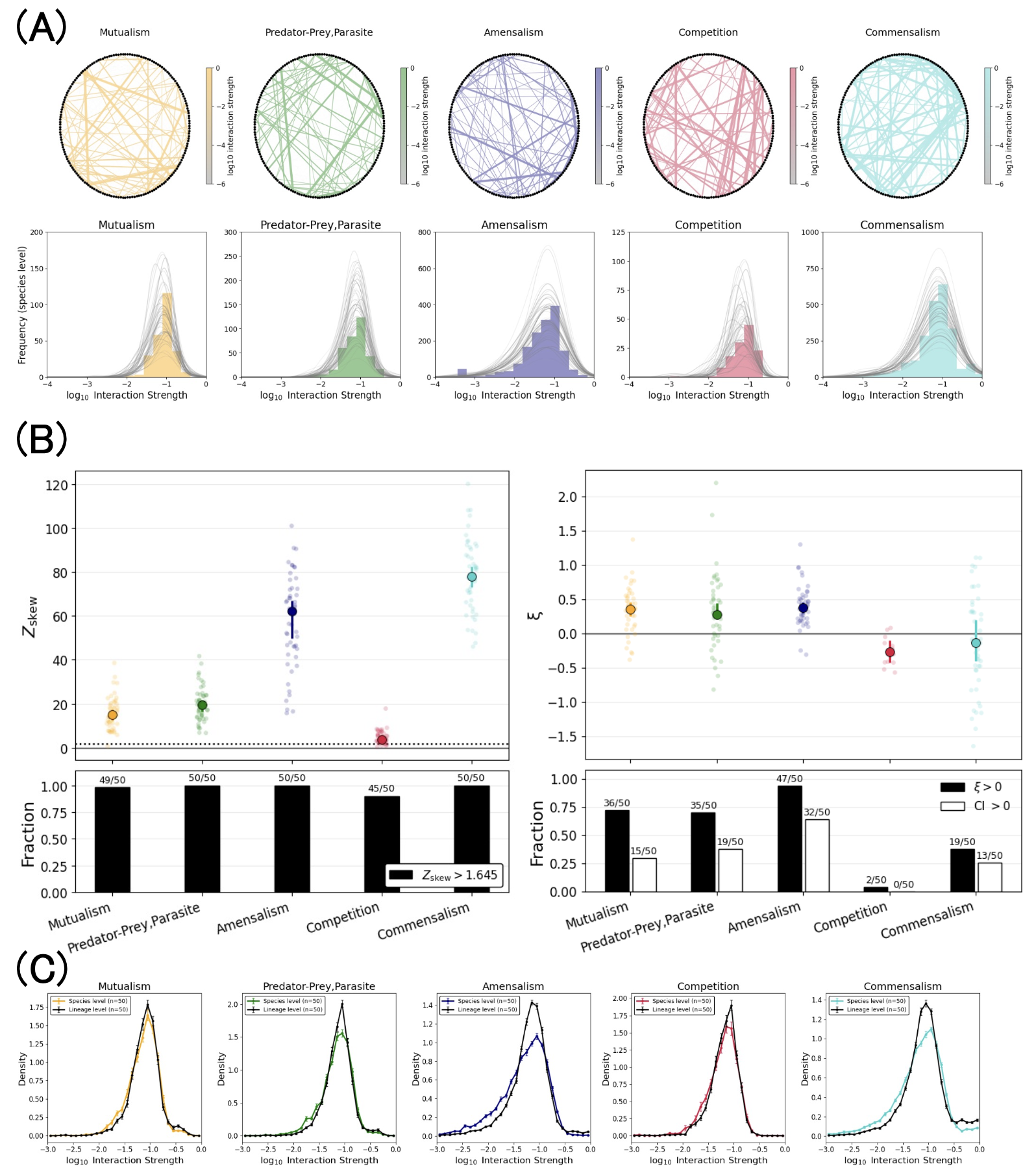}
\caption{\textbf{SWAPS distribution emerges in GLV assembly simulations.} (A) The representative example of interaction network topology (top), frequency distribution of interaction strength (bottom, color barplot) at the 1000 generations with kernel density estimation (KDE, gray) of 50 independent assembly replicates superimposed on a single panel. (B) Black-outlined point indicates the median across replicates with 95\% bootstrap confidence intervals for skewness (left, upper panel) and $\xi$ (right, upper panel) in IS distribution. Small points indicate replicate-level estimates. The lower panel shows the fraction of cases exhibiting significant deviations for each statistic, indicated by black bars (left: $Z_{\mathrm{skew}}>1.645$, right: $\xi > 0$). White bars (right bottom panel) additionally indicate the fraction of cases in which the bootstrap-estimated lower confidence bound of $\xi$ was greater than zero. (C) Probability density distributions of IS at the level of species (each color) and lineage (black), which are averaged across 50 independent assembly replicates. The mean density at each bin (points) and the standard errors (error bars) across replicates are shown. Colors denote interaction types (from left to right: mutualism, predator--prey/parasitism, amensalism, competition, commensalism).}
\label{fig:fig3}
\end{figure}

\begin{figure}[p]
\centering
\includegraphics[width=\textwidth,height=0.68\textheight,keepaspectratio]{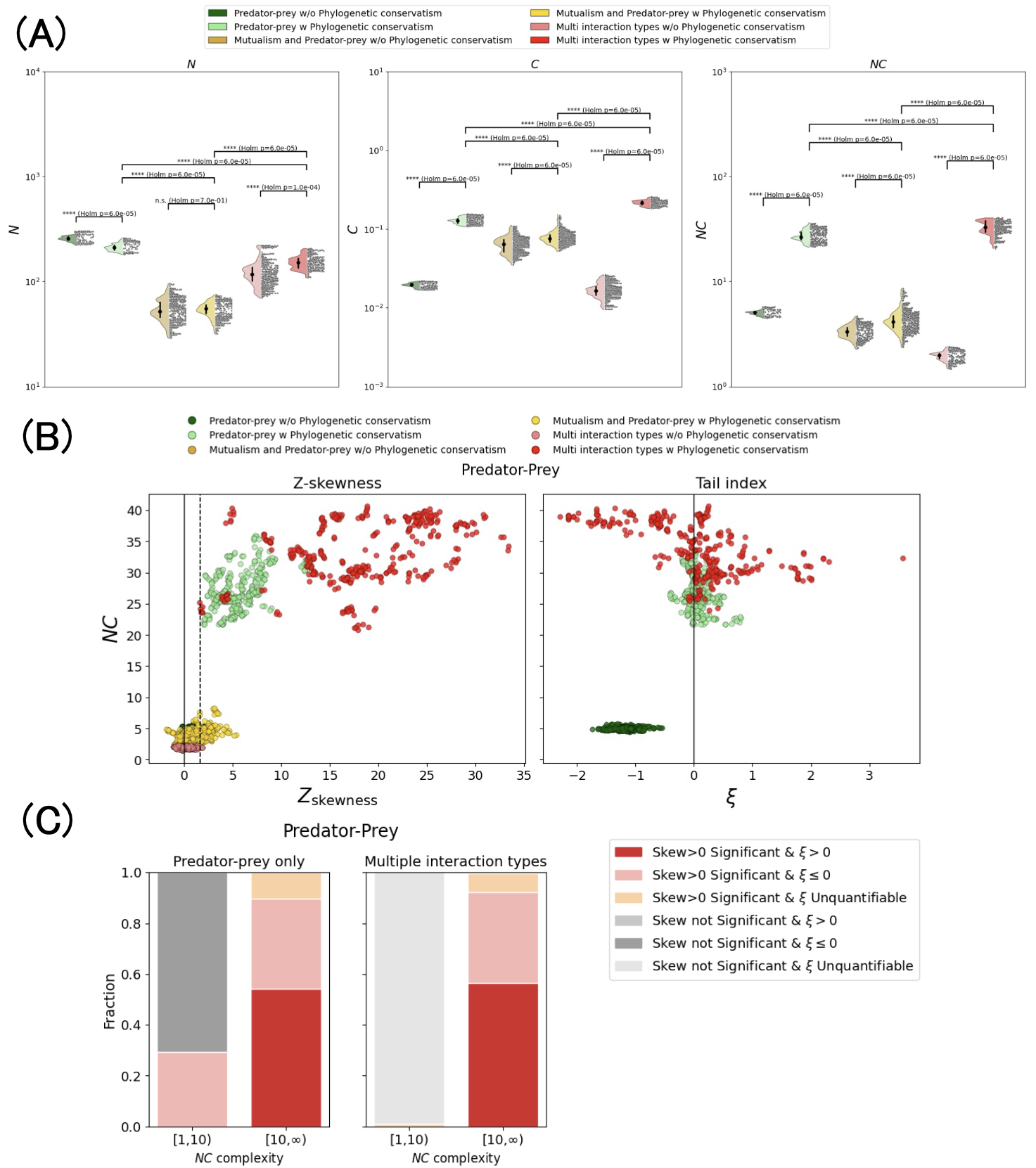}
\caption{\textbf{Community complexity co-emerges with SWAPS distribution.} (A) Species abundance $N$ (left panel), connectance $C$ (center panel), and the product $NC$ (right panel) plotted under six conditions of community assembly simulations (inset); from left to right, violin plots represent predator--prey without (dark green) and with taxonomic conservatism (light green), mutualism and predator--prey without (light brown) and with taxonomic conservatism (ocher), multiple interaction types without (light pink) and with taxonomic conservatism (red). Each scatter plot displays up to 700 randomly sampled points over the last fifty generations per each community assembly. Permutation test on the difference in the means (original scale), with Holm correction for three comparisons (two-sided; 100,000 permutations). (B) Scatter plots of community complexity $NC$ against $Z_{\mathrm{skew}}$ (left panel) and $\xi$ (right panel). Colors indicate the assembly conditions, same as in A. (C) Correlation of skewness and heavy tail with $NC$, binned as $[1,10)$ and $[10,\infty)$, is shown by normalized frequency of the six classes, which are defined by whether skewness is significantly shifted in the positive direction relative to normality ($Z_{\mathrm{skew}}>1.645$, one-sided test) and whether the tail index $\xi$ is positive, non-positive, or unquantifiable. Communities were pooled across distant-only and close/distant immigration simulations, deduplicated after restricting to stable communities, and randomly subsampled to equalize sample sizes among the simulation conditions. Each panel shows the community assembly conditions (left: predator--prey alone, right: multiple interaction types).}
\label{fig:fig4}
\end{figure}

\begin{table}[p]
\centering
\includegraphics[width=\textwidth,height=0.82\textheight,keepaspectratio]{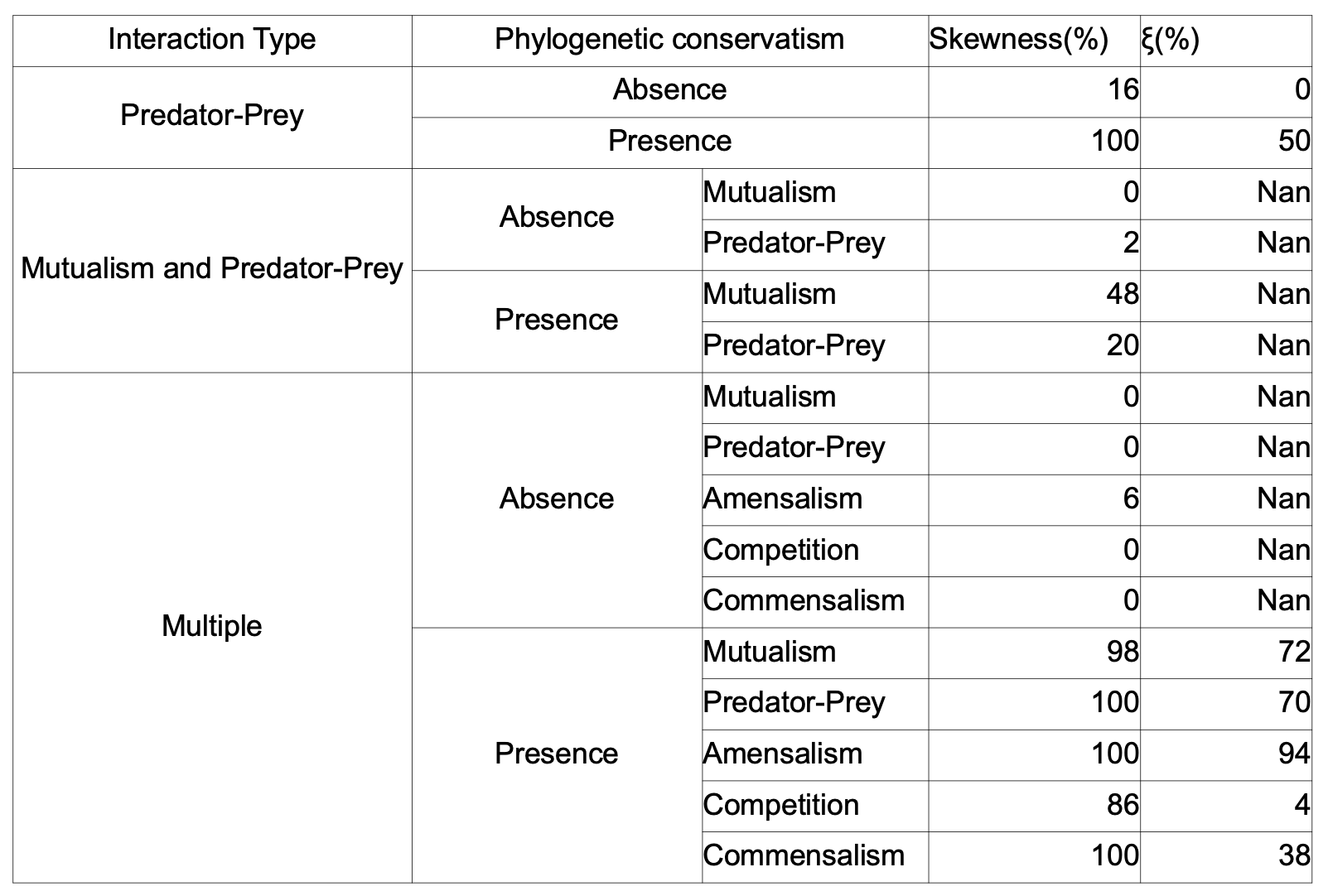}
\caption{\textbf{Frequency of SWAPS emergence out of 50 assembly simulation replicates.} SW and PS interactions were evaluated by a significant deviation of the skewness from normality (D'Agostino's skewness test, $p < 0.01$) and positive Pareto exponent $\xi$, respectively. Community assembly was performed with multiple interaction types and immigration of taxonomically close/distant species.}
\label{tab:table1}
\end{table}

\clearpage
\setcounter{equation}{0}
\renewcommand{\theequation}{S\arabic{equation}}
\renewcommand{\theHequation}{S\arabic{equation}}
\setcounter{figure}{0}
\renewcommand{\thefigure}{S\arabic{figure}}
\renewcommand{\theHfigure}{S\arabic{figure}}
\section*{Supporting Information}

\begin{itemize}[leftmargin=2em]
    \item Supporting Method S1.
    \item Supporting Figures S1--S4.
\end{itemize}

\section*{Supporting Method S1}

\subsection*{Quantification of tail weight based on Peak Over Threshold (POT) theorem}

For each interaction strength distribution, we analyzed the upper tail using a peaks-over-threshold (POT) approach combined with generalized Pareto distribution fitting [58, 59, 60]. This analysis was performed on a one-dimensional positive sample
\begin{linenomath*}
\begin{equation}
    y_1, y_2, \ldots, y_n > 0,
\end{equation}
\end{linenomath*}
where $n$ denotes the number of data points.

To avoid unstable tail inference from a small number of samples, we excluded distributions with insufficient sample size before threshold analysis. Specifically, we required
\begin{linenomath*}
\begin{equation}
    n \geq \max(n_{\min}, 2k_{\min}).
\end{equation}
\end{linenomath*}
We set $n_{\min}=200$ and $k_{\min}=30$; therefore, distributions with $n<n_{\min}$ were not analyzed further.

Next, we defined candidate thresholds from empirical quantiles of the observed distribution. Let $q$ denote a candidate quantile. We considered 25 equally spaced quantiles between 0.80 and 0.99 and defined the corresponding threshold as
\begin{linenomath*}
\begin{equation}
    u(q) = Q_y(q),
\end{equation}
\end{linenomath*}
where $Q_y(q)$ is the empirical $q$-quantile of the interaction strength distribution. For each threshold $u$, we defined exceedances as
\begin{linenomath*}
\begin{equation}
    z_i = y_i - u \quad \text{for all } y_i > u,
\end{equation}
\end{linenomath*}
and the number of exceedances as
\begin{linenomath*}
\begin{equation}
    k = \#\{y_i: y_i > u\}.
\end{equation}
\end{linenomath*}
The exceedance fraction $k/n$ was also recorded.

Thresholds that are too low may include substantial non-tail data, whereas thresholds that are too high can lead to unstable inference because of small exceedance counts. We therefore determined admissible thresholds using both absolute and relative exceedance counts. For each distribution, the effective lower and upper bounds on $k$ were defined as
\begin{linenomath*}
\begin{align}
    k_{\mathrm{eff},\min} &= \max(k_{\min}, n f_{\min}), \\
    k_{\mathrm{eff},\max} &= \min(k_{\max}, n f_{\max}).
\end{align}
\end{linenomath*}
If necessary, the upper bound was adjusted so that
\begin{linenomath*}
\begin{equation}
    k_{\mathrm{eff},\max} \geq k_{\mathrm{eff},\min}.
\end{equation}
\end{linenomath*}
In this analysis, we set $k_{\max}=200$, $f_{\min}=0.05$, and $f_{\max}=0.20$. Thus, candidate thresholds were restricted to those yielding at least 30 exceedances and at least 5\% of the total sample, but at most 200 exceedances and at most 20\% of the total sample. Threshold-specific fitting was performed only for candidates with
\begin{linenomath*}
\begin{equation}
    k \geq k_{\mathrm{eff},\min},
\end{equation}
\end{linenomath*}
and final threshold selection was restricted to candidates satisfying
\begin{linenomath*}
\begin{equation}
    k_{\mathrm{eff},\min} \leq k \leq k_{\mathrm{eff},\max}.
\end{equation}
\end{linenomath*}

At each candidate threshold, the exceedance distribution was fitted by a generalized Pareto distribution with the location fixed at zero. Writing the shape parameter as $\xi$ and the scale parameter as $\beta$, the density for $\xi 
eq 0$ is
\begin{linenomath*}
\begin{equation}
    f(z) = \frac{1}{\beta}\left(1 + \frac{\xi z}{\beta}\right)^{-1/\xi - 1}
    \quad \left(z \geq 0,\; 1 + \frac{\xi z}{\beta} > 0\right).
\end{equation}
\end{linenomath*}
When $\xi=0$, this distribution reduces to the exponential distribution. Parameters $\xi$ and $\beta$ were estimated by maximum likelihood with the location parameter fixed at zero.

For the fitted generalized Pareto distribution, the log-likelihood was evaluated as
\begin{linenomath*}
\begin{equation}
    \ell_{\mathrm{GPD}} = -k\log\beta - \left(\frac{1}{\xi}+1\right)\sum_{i=1}^{k}\log\left(1 + \frac{\xi z_i}{\beta}\right)
    \quad (\xi 
eq 0),
\end{equation}
\end{linenomath*}
provided that $\beta>0$ and the support constraint was satisfied for all exceedances. When $\xi=0$, the corresponding exponential log-likelihood is
\begin{linenomath*}
\begin{equation}
    \ell_{\mathrm{Exp}} = -k\log\beta - \sum_{i=1}^{k}\frac{z_i}{\beta}.
\end{equation}
\end{linenomath*}

As a reference comparison, an exponential model was also fitted to the same exceedance sample. Under this model, the scale parameter was estimated by the sample mean of the exceedances,
\begin{linenomath*}
\begin{equation}
    \beta_{\mathrm{Exp}} = \frac{1}{k}\sum_{i=1}^{k} z_i.
\end{equation}
\end{linenomath*}
To compare the generalized Pareto and exponential fits, we computed the likelihood-ratio statistic
\begin{linenomath*}
\begin{equation}
    LR = 2(\ell_{\mathrm{GPD}} - \ell_{\mathrm{Exp}}),
\end{equation}
\end{linenomath*}
and an approximate $p$-value based on the chi-squared distribution with one degree of freedom,
\begin{linenomath*}
\begin{equation}
    p_{LR} = 1 - F_{\chi^2_1}(LR).
\end{equation}
\end{linenomath*}
This quantity was used only as a descriptive diagnostic and was not used for threshold selection or final heavy-tail classification.

To quantify uncertainty in the estimated shape parameter $\xi$, we performed nonparametric bootstrap resampling for each candidate threshold. Let $z_1,z_2,\ldots,z_k$ denote the exceedances at a given threshold. In each bootstrap replicate, we resampled $k$ exceedances with replacement from the observed exceedance sample, refitted the generalized Pareto distribution, and stored the bootstrap estimate of the shape parameter, $\xi^{(b)}$. This procedure was repeated $B=500$ times. When a sufficient number of finite bootstrap estimates was obtained, the 95\% bootstrap confidence interval (CI) for $\xi$ was defined by the empirical 2.5th and 97.5th percentiles of the bootstrap distribution:
\begin{linenomath*}
\begin{equation}
    [\xi_{\mathrm{low}}, \xi_{\mathrm{high}}].
\end{equation}
\end{linenomath*}
If the exceedance count was too small or too few finite bootstrap estimates were obtained, the CI was treated as undefined.

For each candidate threshold, we recorded the quantile $q$, threshold $u$, total sample size $n$, exceedance count $k$, exceedance fraction $k/n$, maximum-likelihood estimates of $\xi$ and $\beta$, bootstrap confidence limits $\xi_{\mathrm{low}}$ and $\xi_{\mathrm{high}}$, generalized Pareto and exponential log-likelihoods, likelihood-ratio statistic, and approximate likelihood-ratio $p$-value.

We selected the final threshold without using the estimated value of $\xi$ or its CI, in order to avoid circularity between threshold selection and tail-heaviness assessment. Among the admissible candidate thresholds satisfying
\begin{linenomath*}
\begin{equation}
    k_{\mathrm{eff},\min} \leq k \leq k_{\mathrm{eff},\max},
\end{equation}
\end{linenomath*}
we computed the distance between the observed exceedance count and a prespecified target count $k_{\mathrm{target}}$,
\begin{linenomath*}
\begin{equation}
    d_k = |k - k_{\mathrm{target}}|.
\end{equation}
\end{linenomath*}
In this study, we primarily used $k_{\mathrm{target}}=50$ and additionally confirmed that the results were qualitatively robust to alternative choices of $k_{\mathrm{target}}=30$, 80, and 100, indicating that the inference did not strongly depend on this parameter (Fig.~S1B). The final threshold $u^*$ was chosen as the candidate minimizing $d_k$; in case of ties, the candidate with the larger quantile $q$ was selected. Thus, threshold selection depended only on exceedance-count criteria and favored the highest admissible threshold among equally suitable candidates.

For the selected threshold $u^*$, we reported the corresponding quantile $q^*$, exceedance count $k^*$, exceedance fraction $k^*/n$, shape estimate $\xi^*$, its bootstrap CI, and the approximate likelihood-ratio $p$-value. A distribution was classified as having a heavy upper tail when the lower bound of the bootstrap CI for the shape parameter at the selected threshold exceeded zero, that is,
\begin{linenomath*}
\begin{equation}
    \xi^*_{\mathrm{low}} > 0.
\end{equation}
\end{linenomath*}
Accordingly, heavy-tail classification was based not merely on $\xi^*>0$, but on whether the bootstrap CI lay entirely above zero.

For visualization, we generated a diagnostic plot for each dataset showing the estimated shape parameter $\xi$ as a function of the threshold quantile $q$, together with the bootstrap CI at each candidate threshold, a horizontal reference line at $\xi=0$, and a vertical line indicating the selected threshold $q^*$. For cross-dataset comparison, we additionally summarized the selected threshold estimates $\xi^*$ and their CIs across datasets.

\clearpage

\begin{figure}[p]
    \centering
    \supportinggraphic[width=\textwidth,height=0.72\textheight,keepaspectratio]{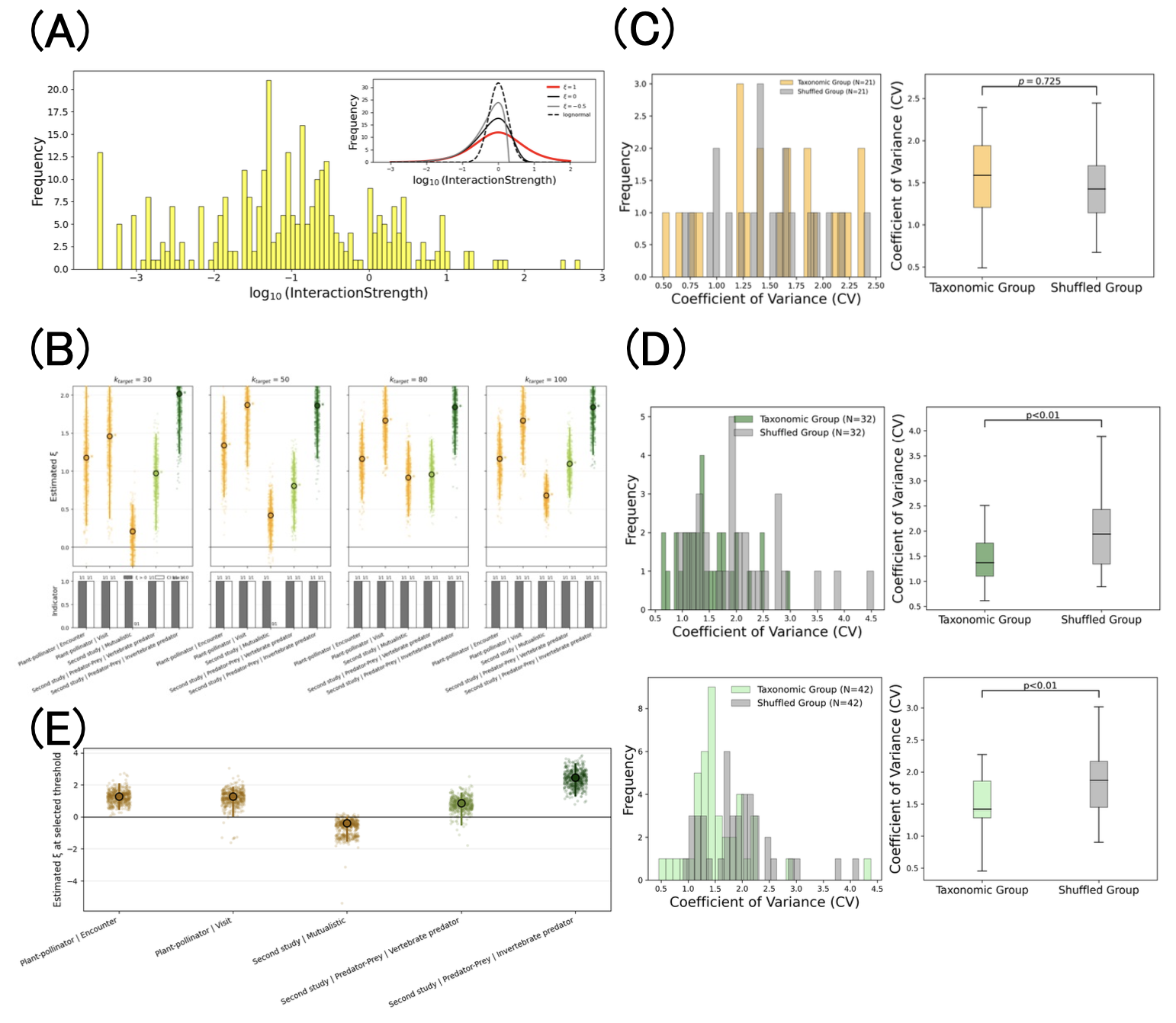}
    \caption{\textbf{Additional empirical data analysis, related to Figures 1 and 2.}
    (A) Frequency distributions of interaction strength quantified by encounter-based approaches in empirical data from a plant--pollinator network [49]. (B) The tail shape parameter $\xi$ was estimated using a peaks-over-threshold (POT) approach, where the key design parameter $k_{\mathrm{target}}$, the target number of exceedances used to choose the final threshold, was set to different values (30, 50, 80, and 100) in each panel. The sign of $\xi$ was insensitive to the parameter value. (C--D) Coefficient of variation (CV) of the averaged interaction strength (IS) within taxonomic groups with four or more mutualistic IS values (orange) and shuffled groups (grey) in a plant--pollinator network (C) and agroecosystem (D, upper: invertebrate predator; D, lower: vertebrate predator), shown as a histogram (left) and boxplot (right). (E) Tail shape parameter $\xi$ at the family level of the interaction strength distribution.}
    \label{fig:s1}
\end{figure}

\begin{figure}[p]
    \centering
    \supportinggraphic[width=\textwidth,height=0.64\textheight,keepaspectratio]{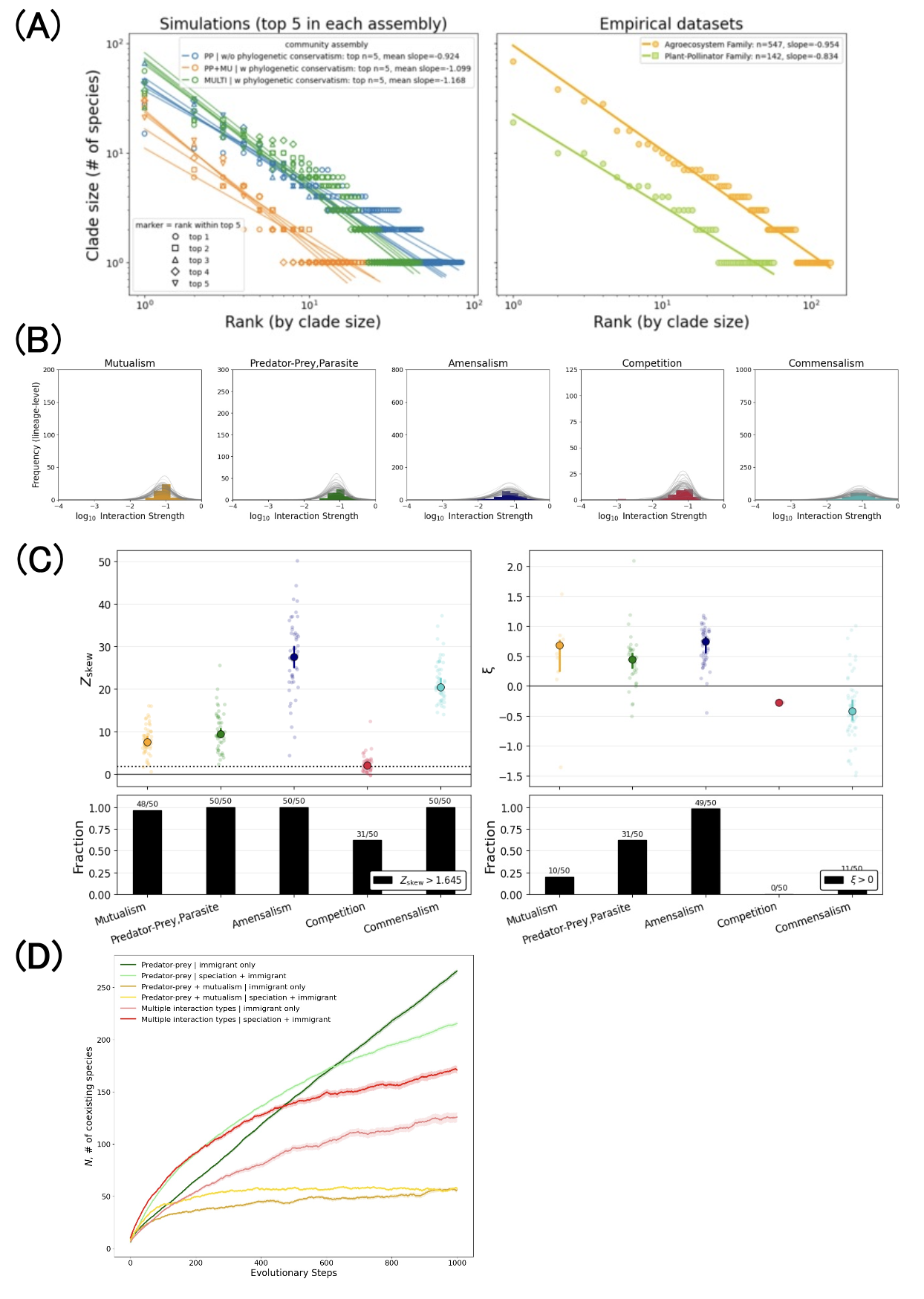}
    \caption{\textbf{Distributions of clade rank size and interaction strength at the lineage level, related to Figure 3.}
    (A) Clade size rank distributions are shown for simulated communities at the lineage level (left) and empirical communities at the family level (right). Colors indicate assembly conditions (left inset: predator--prey only, blue; mutualism and predator--prey, orange; and multiple interaction types, green) and datasets (right inset: plant--pollinator network, orange [49]; and agroecosystem, light green [50]), respectively. Colored symbols and lines represent clade size within a community and the linear regression on a double-logarithmic scale. The five communities with the highest species richness after 1,000 generations, which are used in Figure 3, are shown for each assembly condition. (B) Frequency distributions of interaction strength (IS) at the lineage level after 1,000 generations. Colored bars indicate the focal lineage, whereas kernel density estimates (gray) indicate the IS distributions in 50 independent assembly replicates. (C) Skewness and tail-shape analyses of lineage-level IS distributions for each interaction type shown by color (from left to right: mutualism, predator--prey/parasitism, amensalism, competition, and commensalism). Upper panels show the skewness indicator $Z_{\mathrm{skew}}$ (left) and tail-shape parameter $\xi$ (right). Black-edged points indicate the median across lineages, whereas other points indicate bootstrap estimates. Lower panels show the fraction of cases meeting each criterion, indicated by black bars: $Z_{\mathrm{skew}}>1.645$ (left) and $\xi>0$ (right), and the fraction meeting the positive lower confidence-bound criterion of the bootstrap estimation. (D) Dynamics of species richness $N$ during community assembly simulations, colored by assembly condition (inset) in the same manner as Figure 4. The mean of 50 replicates (thick line) and standard error (error bar) are shown.}
    \label{fig:s2}
\end{figure}

\begin{figure}[p]
    \centering
    \supportinggraphic[width=\textwidth,height=0.68\textheight,keepaspectratio]{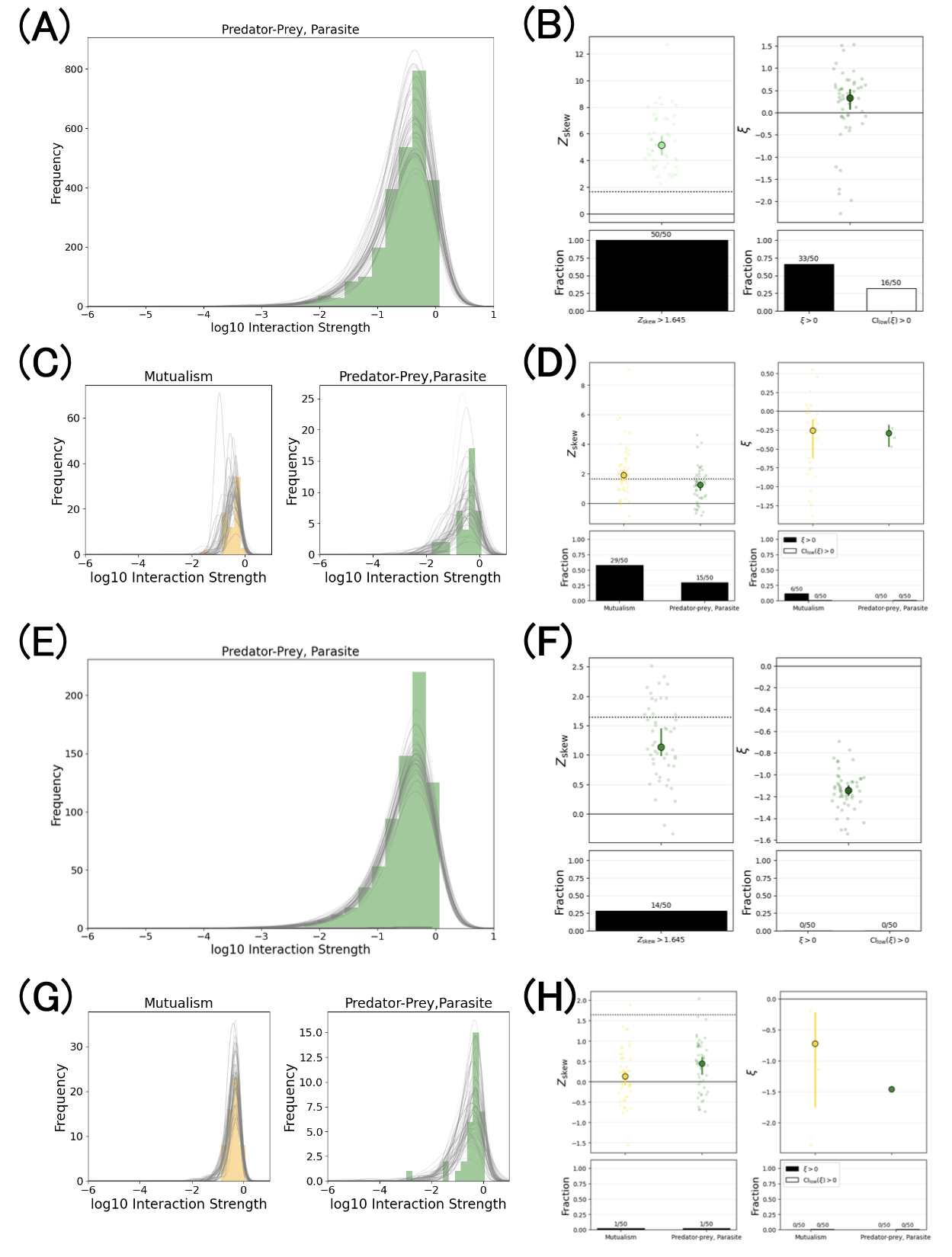}
    \caption{\textbf{Lineage-level statistics of interaction strength across assembly conditions, related to Figure 3.}
    (A, C, E, G, I) Frequency distributions of interaction strength (IS) at the lineage level and (B, D, F, H, J) skewness and tail-shape analyses of lineage-level IS distributions after 1,000 generations for each assembly condition: predator--prey alone with taxonomic conservatism (A, B), mutualism and predator--prey with taxonomic conservatism (C, D), predator--prey alone without taxonomic conservatism (E, F), mutualism and predator--prey without taxonomic conservatism (G, H), and multiple interaction types without taxonomic conservatism (I, J). Colors represent interaction types. Upper panels of B, D, F, H, and J show the median (black outlined circle) of $Z_{\mathrm{skew}}$ (left) and $\xi$ (right) across lineages, and bootstrap estimates (points). Lower panels show the proportion of cases meeting each criterion, indicated by black bars: $Z_{\mathrm{skew}}>1.645$ in the left panel and $\xi>0$ in the right panel. For the $\xi$ analysis, white bars additionally indicate the proportion of cases in which the bootstrap-estimated lower confidence bound of $\xi$ was greater than zero.}
    \label{fig:s3}
\end{figure}

\begin{figure}[p]
    \centering
    \supportinggraphic[width=\textwidth,height=0.80\textheight,keepaspectratio]{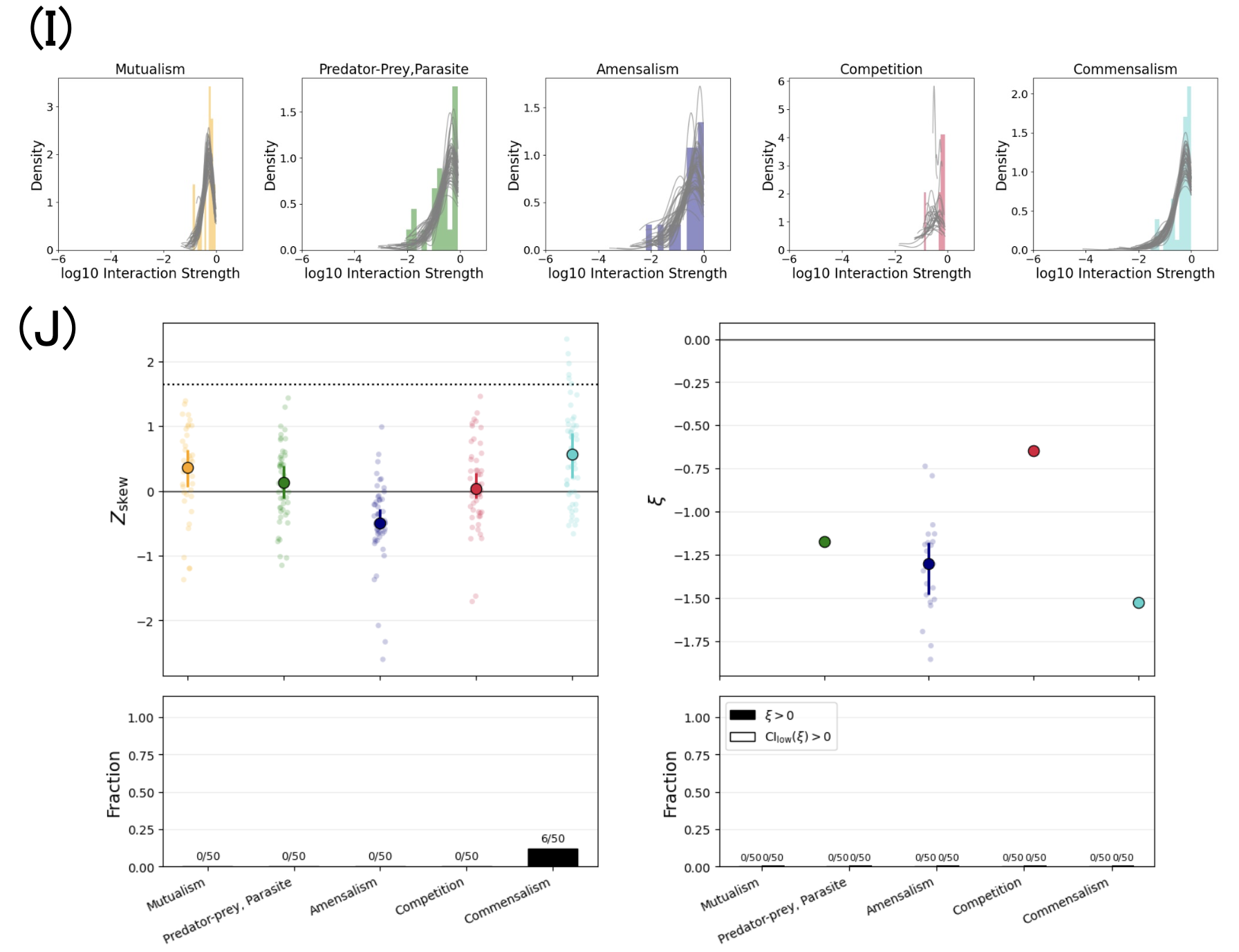}
    \caption*{\textbf{Figure S3 continued.} Lineage-level statistics of interaction strength across assembly conditions.}
\end{figure}

\begin{figure}[p]
    \centering
    \supportinggraphic[width=\textwidth,height=0.70\textheight,keepaspectratio]{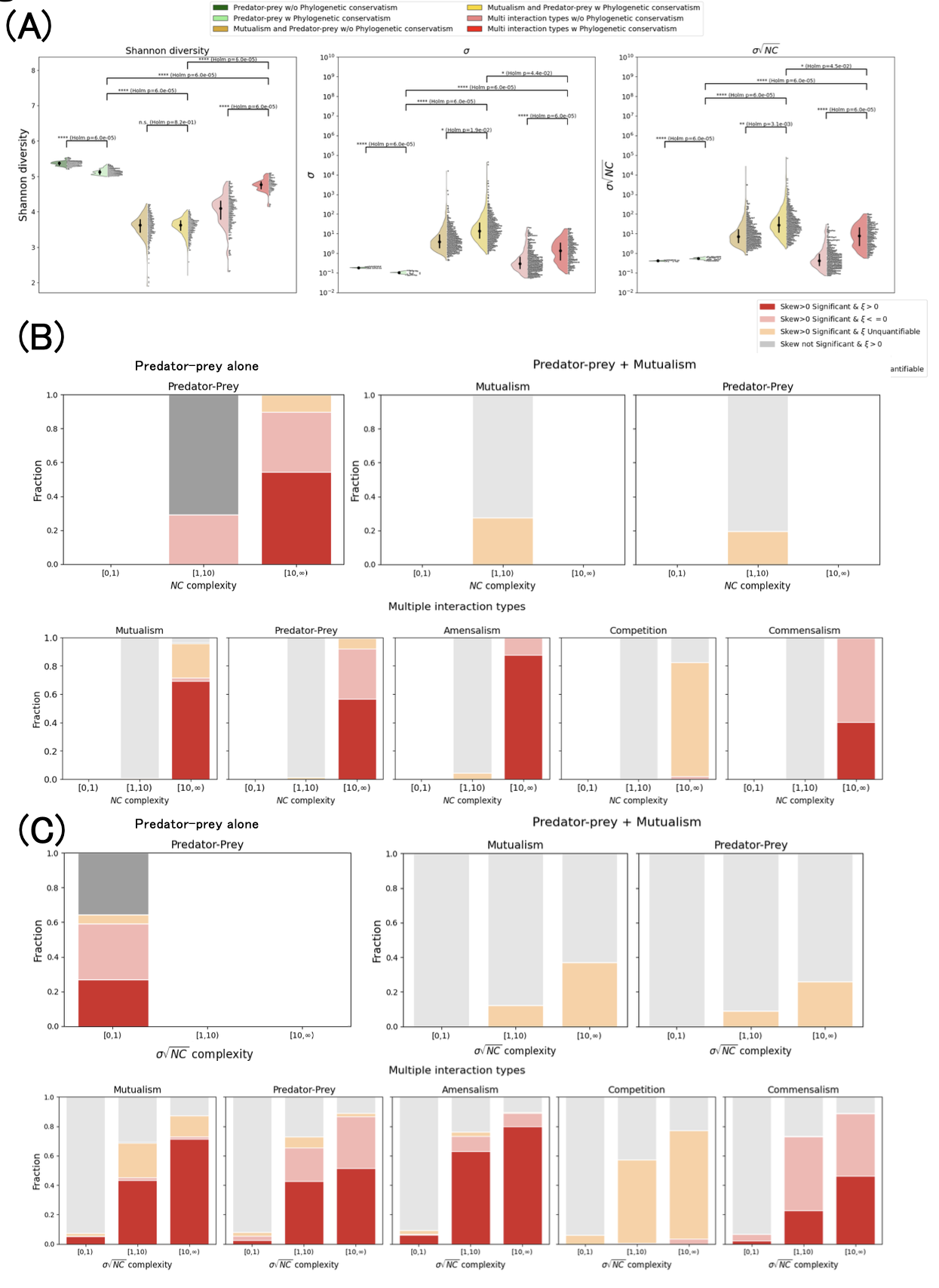}
    \caption{\textbf{Relationship between community signatures and and SWAPS emergence, related to Figure 4.}
    (A) Shannon diversity (left panel), interaction strength variability $\sigma$ (center panel), and May's complexity $\sigma\sqrt{NC}$ (right panel) plotted under six conditions of community assembly simulations (inset). Datasets of assembly simulations, colors and arrangement are the same as those in Fig. 4A. Permutation test on the difference in the means (original scale), with Holm correction for three comparisons (two-sided; 100,000 permutations). (B-C) Correlation of skewness and heavy tail with $NC$ (B) and $\sigma\sqrt{NC}$ (C), binned as $[1,10)$ and $[10,\infty)$, is shown by normalized frequency of the six classes, which are defined by whether skewness is significantly shifted in the positive direction relative to normality ($Z_{\mathrm{skew}}>1.645$, one-sided test) and whether the tail index $\xi$ is positive, non-positive, or unquantifiable. Communities were pooled across distant-only and close/distant immigration simulations, deduplicated after restricting to stable communities, and randomly subsampled to equalize sample sizes among the simulation conditions. Each panel shows the community assembly conditions (upper left: predator-prey alone, upper right: mutualism and predator-prey , bottom: multiple interaction types).}
    \label{fig:s4}
\end{figure}

\end{document}